\documentclass[]{article}
\usepackage{amsmath, amscd}
\usepackage[pdftex]{graphicx} 
\usepackage{graphicx}
\usepackage{xcolor}
\usepackage{verbatim}

\usepackage{rotating}

\usepackage{amsmath}
\usepackage{datetime}
\definecolor{Black}{rgb}{0,0,0}
\definecolor{Red}{rgb}{0.9,0.2,0.1}

\newcommand{\ODeltaEs}{O$\mathrm{\Delta}$Es}

\newcommand{\PDeltaE}{P$\mathrm{\Delta}$E}
\newcommand{\PDeltaEs}{P$\mathrm{\Delta}$Es}
\newcommand{\dsp}{\displaystyle}
\newcommand{\tg}{ {\tilde{g}} }
\newcommand{\tf}{ {\tilde{f}} }
\newcommand{\tF}{ {\tilde{F}} }
\newcommand{\tG}{ {\tilde{G}} }

\begin{document}
%
%
%
%
\title{Symbolic Computation of Lax Pairs of Partial Difference Equations 
using Consistency Around the Cube}
%
%
%
%
\author{T.\ Bridgman$^{a}
        \thanks{Corresponding author. Email: tbridgma@mines.edu }\;$,
        W.\ Hereman$^{a}$,
        G.\ R.\ W.\ Quispel$^{b}$,  
        P.\ H.\ van der Kamp$^{b}$ \\  \\
        $^a$ Department of Applied Mathematics and Statistics, \\
        Colorado School of Mines, Golden, CO 80401-1887, U.S.A. \\  
        $^b$ Department of Mathematics and Statistics,  \\
        La Trobe University, Melbourne, Victoria 3086, Australia 
        }
%

\date{Received: March 8, 2012 / Accepted: May ??, 2012}

\maketitle
%
\vskip 1pt
\noindent
\centerline{Communicated by Elizabeth Mansfield}
\vskip 12pt
\noindent
\centerline{{\bf Dedication}}
\vskip 1pt
\noindent 
At the occasion of his 60th birthday, we like to dedicate this paper to 
Peter Olver whose work has inspired us throughout our careers. 
%
\vskip 10pt
\noindent
\begin{abstract}
A three-step method due to Nijhoff and Bobenko \& Suris to derive a Lax pair 
for scalar partial difference equations $({\rm P}\Delta{\rm Es})$ is reviewed.
The method assumes that the ${\rm P}\Delta{\rm Es}$ are defined on a 
quadrilateral, and consistent around the cube.
Next, the method is extended to systems of ${\rm P}\Delta{\rm Es}$ 
where one has to carefully account for equations defined on edges of the 
quadrilateral.
Lax pairs are presented for scalar integrable ${\rm P}\Delta{\rm Es}$ 
classified by Adler, Bobenko, and Suris 
and systems of ${\rm P}\Delta{\rm Es}$ including the integrable 2-component 
potential Korteweg-de Vries lattice system, as well as nonlinear 
Schr\"odinger and Boussinesq-type lattice systems. 
Previously unknown Lax pairs are presented for ${\rm P}\Delta{\rm Es}$
recently derived by Hietarinta (J.\ Phys.\ A: Math.\ Theor., 44 (2011) 
Art.\ No.\ 165204).
The method is algorithmic and is being implemented in {\sc Mathematica}. 
\end{abstract}

%
%
%
%

%
\section{Introduction}
\label{sec:intro}
The original Lax pair \cite{Lax1968} was a duo of commuting linear 
differential operators representing the integrable Korteweg-de Vries (KdV)
equation. 
Lax's idea was to replace a nonlinear partial differential equation (PDE), 
such as the KdV equation, by a pair of {\em linear} PDEs of high-order 
(in an auxiliary eigenfunction) whose compatibility requires that the 
{\em nonlinear} PDE holds.
One can write these high-order linear PDEs as a system of PDEs of first order;
hence, replacing the Lax operators with a pair of matrices.
The Lax equation to be satisfied by these matrices is commonly referred 
to as the zero-curvature representation \cite{ZakharovShabat1974} 
of the nonlinear PDE. 
The discovery of Lax pairs was crucial for the further development of the 
inverse scattering transform (IST) method which had been introduced 
in \cite{GGKM1967}. 

For partial difference equations (P$\Delta$Es) Lax pairs first appeared 
in the work of Ablowitz and Ladik \cite{AblowitzLadik1976,AblowitzLadik1977}, 
and subsequently in \cite{NijhoffQuispelCapel1983} for other equations. 
The fundamental characterization of {\em integrable} P$\Delta$Es as being 
multi-dimensionally consistent \cite{BobenkoSuris2002,Nijhoff2002} 
is intimately related to the existence of a Lax pair.

Lax pairs for P$\Delta$Es are not only crucial for applying the IST, 
they can be used to construct integrals for mappings and correspondences 
obtained as periodic reductions, using the so-called staircase method. 
This method was developed in \cite{Papageorgiouetal1990} and extended 
in \cite{Quispeletal1991} to cover more general reductions. 
Essential to the staircase method is the construction of a product of Lax 
matrices (the {\em monodromy matrix}) whose characteristic polynomial is an 
invariant of the evolution.
In fact, the monodromy matrix can be interpreted as one of the Lax matrices 
for the reduced mapping 
\cite{QuispelCapelRoberts2005,Rojas2009,RojasvanderKampQuispel2007}.
Through expansion of the characteristic equation of the monodromy matrix 
in the spectral parameters a number of functionally independent invariants 
can be obtained. 
A recent investigation \cite{vanderKampQuispel2010} supports the idea that 
the staircase method provides sufficiently many integrals for the periodic 
reductions to be completely integrable (in the sense of Liouville-Arnold).

Finding a Lax pair for a given nonlinear equation, whether continuous or
discrete, is generally a difficult task. 
For PDEs the theory of pseudo potentials \cite{WahlquistEstabrook1975} 
might lead to a Lax pair, but it only works in certain cases. 
The most powerful method to find Lax pairs is the dressing method 
developed by Zakharov and Shabat in 1974 (see, e.g., 
\cite{ZakharovManakovNovikovPitaevskii1994}).
Building on the key idea of the dressing method, there exists a 
straightforward, algorithmic approach to derive a Lax pair 
\cite{BobenkoSuris2002,Nijhoff2002} for scalar P$\Delta$Es that are 
{\em consistent around the cube} (CAC).
That approach is reviewed in Section \ref{sec:scCAC}. 
In Section \ref{sec:systemPDE}, it is applied to systems of lattice equations, 
as was done in \cite{TongasNijhoff2005,Walker2001} for the case of the 
Boussinesq system.

We are currently developing {\sc Mathematica} software for the symbolic 
computation of Lax pairs for lattice equations \cite{Bridgman2012,Hereman2009}.
Section \ref{sec:implement} outlines the implementation strategy for the 
verification of the CAC property and the computation (and subsequent 
verification) of the Lax pair.
%
With the exception of the $Q_4$ equation whose Lax pair was given in 
\cite{Nijhoff2002}, the software has been used to produce Lax pairs of the 
ABS equations \cite{AdlerBobenkoSuris2002} and the $(\alpha,\beta$)-equation. 
The latter is also known as the NQC equation after Nijhoff, Quispel, and 
Capel \cite{NijhoffQuispelCapel1983} and its Lax pair was first reported 
in \cite{Tran2007}.

With respect to lattice systems, we computed Lax pairs of the Boussinesq 
and Toda-modified Boussinesq systems \cite{Nijhoffetal1992}, 
as well as the Schwarzian Boussinesq \cite{Nijhoff1996} and modified 
Boussinesq \cite{XenitidisNijhoff2011} systems. 
Using the code, we also computed Lax pairs for the 2-component potential KdV 
and nonlinear Schr\"odinger systems \cite{Mikhailov2009,Xenitidis2009}. 
Details of the calculations, and alternative Lax pairs, are given in 
Section \ref{sec:results}.
We obtained new Lax pairs for the 2- and 3-component Hietarinta systems 
\cite{Hietarinta2011}. 
In contrast to the $4\times 4$ Lax matrices for the Hietarinta systems 
\cite{Hietarinta2011} obtained (independently) in 
\cite{ZhangZhaoNijhoff2012}, the Lax matrices presented in this paper 
are $3\times 3$ matrices.
\section{Scalar partial difference equations}
\label{sec:scalarPDE}
\subsection{Consistency around the cube for scalar \PDeltaEs}
\label{sec:scCAC}
The concept of multi-dimensional consistency was introduced independently 
in \cite{BobenkoSuris2002,NijhoffWalker2001}. 
The key idea is to embed the equation consistently into a multi-dimensional 
lattice by imposing copies of the same equation, albeit with different 
lattice parameters in different directions. 
The consistency for embedding a 2-dimensional lattice equation, 
defined on an elementary quadrilateral, into a 3-dimensional lattice on a 
cube is commonly referred to as consistency around the cube (CAC). 
%
For multi-affine nonlinear \PDeltaEs\, with the CAC property there is an 
algorithmic way of deriving a Lax pair.
\begin{figure*}[h]
 \centering
  \includegraphics[scale=1]{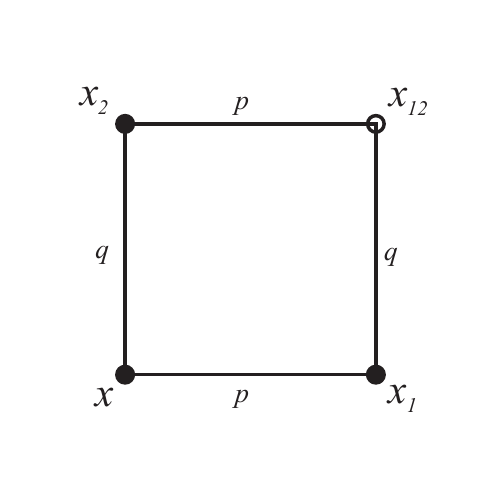}
  \caption{The \PDeltaE\, is defined on the simplest quadrilateral
   (a square).}
  \label{fig:square}       
\end{figure*}
\newline
\indent
In this paper we consider \PDeltaEs,
\begin{equation}
\label{E:PDeltaE_F}
{\cal F}(x, x_1, x_2, x_{12}; p, q) = 0,
\end{equation}
which are defined on a 2-dimensional quad-graph as shown in 
Figure~\ref{fig:square}.
The field variable $x = x_{n,m}$ depends on lattice variables $n$ and $m.$
A shift of $x$ in the horizontal direction (the $1-$direction) 
is denoted by $x_1 \equiv x_{n+1,m}.$ 
A shift in the vertical or $2-$direction by $x_2 \equiv x_{n,m+1}$ and a 
shift in both directions by $x_{12} \equiv x_{n+1,m+1}.$
Furthermore, ${\cal F}$ depends on the lattice parameters $p$ and $q$ 
which correspond to the edges of the quadrilateral.
Alternate notations are used in the literature.
For instance, many authors denote $(x,x_1,x_2,x_{12})$ by 
$(x, {\tilde{x}},{\hat{x}},{\hat{\tilde{x}}})$ while others use 
$(x_{00}, x_{10}, x_{01}, x_{11}).$

In this paper, we assume that the initial values (indicated by solid circles) 
for $x, x_1$ and $x_2$ can be specified and that the value of $x_{12}$
(indicated by an open circle) can be uniquely determined by
\eqref{E:PDeltaE_F}. 
%
To have single-valued maps, we assume that ${\cal F}$ is multi-affine 
\cite{BobenkoSuris2002}, which is sometimes called multi-linear. 
Atkinson \cite{Atkinson2011} and Atkinson \& Nieszporksi 
\cite{AtkionsonNieszporski2012} have 
recently given examples of \PDeltaEs\ that are multi-{\it quadratic} and 
multi-dimensionally consistent.  

In the simplest case, ${\cal F}$ is a {\em scalar} relation between values 
of a single dependent variable $x$ and its shifts 
(located at the vertices of an elementary square). 
Nonlinear lattice equations of type \eqref{E:PDeltaE_F} arise, for
example, as the permutability condition for B\"acklund transformations
associated with integrable partial differential equations 
(PDEs). 

In more complicated cases, ${\cal F}$ is a nonlinear {\em vector}
function of the vector ${\bf x}$ with several components. 
In that case, \eqref{E:PDeltaE_F} represents a system of \PDeltaEs.
These systems are called multi-component lattice equations.
In such systems some equations might only be defined on the edges 
of the square while others are defined on the whole square. 
The vector case will be considered in Section \ref{sec:systemPDE}.
\noindent
\begin{figure*}[h]
  \centering
  \includegraphics[scale=1.1]{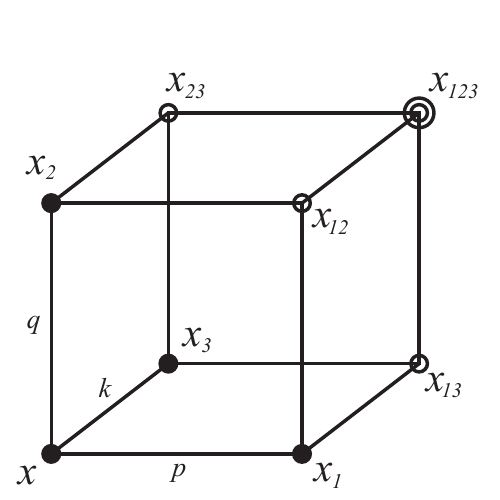}
  \caption{The \PDeltaE\, holds on each face of the cube.}
  \label{fig:cube}       
\end{figure*}
\vskip 0.001pt
\indent
To arrive at a cube, the planar quadrilateral is extended into the third 
dimension as shown in Figure~\ref{fig:cube}, where parameter $k$ is 
the lattice parameter in the third direction. 
%
Although not explicitly shown in Figure~\ref{fig:cube}, all parallel edges
carry the same lattice parameters.

A key assumption is that the original equation(s) holds on all faces of 
the cube.
These equations can therefore be generated by changes of variables and 
parameters, or shifts of the original \PDeltaE. 
On the cube, they can be visualized as either translations, or rotations 
of the faces. 
For example, the equation on the {\rm{left}} face can be obtained via
a rotation of the {\em{front}} face along the vertical axis connecting
$x$ and $x_2.$
This amounts to applying to \eqref{E:PDeltaE_F} the substitutions 
\begin{equation}
\label{E:FtoL}
 x_1 \rightarrow x_3, \;
 x_{12} \rightarrow x_{23}, \; \text{ and } \;
 p \rightarrow k,
\end{equation}
yielding ${\cal F}(x, x_3, x_2, x_{23}; k, q) = 0.$ 
The equation on the {\em{back}} face of the cube can be generated via 
a shift of \eqref{E:PDeltaE_F} in the third direction, letting
\begin{equation}
\label{E:FtoBK}
 x \rightarrow x_3, \;
 x_1 \rightarrow x_{13}, \;
 x_2 \rightarrow x_{23}, \;
 \text{ and } \; x_{12} \rightarrow x_{123}, 
\end{equation}
which yields 
${\cal F}(x_3, x_{13}, x_{23}, x_{123}; p, q) = 0.$

The equations on the {\em{back}}, {\em{right}}, and {\em{top}} faces of 
the cube all involve the unknown $x_{123}$ (indicated by the double
open circle). 
Solving them yields three expressions for $x_{123}.$ 
Consistency around the cube of the \PDeltaE\, requires that one can
uniquely determine $x_{123}$ and that all three expressions coincide.
As discussed in \cite{SurisBobenko2008}, this three-dimensional consistency 
establishes integrability.

The consistency property does not depend on the actual mappings used to 
generate the \PDeltaEs\, on the various faces of the cube. 
Mappings such as \eqref{E:FtoL} and \eqref{E:FtoBK}, which express the 
symmetries of the \PDeltaEs\, are merely a tool for generating the 
needed \PDeltaEs\, quickly.
%
\vskip 3pt
\noindent
{\bf Example 1}:
Consider the lattice modified KdV (mKdV) equation \cite{SurisBobenko2008}
(also classified as 
$H_3$ with $\delta = 0$ as listed in Table~\ref{table:scalarA}), 
\begin{equation}
\label{E:H3d0_F}
p (x x_1 + x_2 x_{12}) - q(x x_2 + x_1 x_{12}) = 0.
\end{equation}
This equation is defined on the {\em{front}} face of the cube.
To verify CAC, variations of the original \PDeltaE\, 
on the {\em{left}} and {\em{bottom}} faces of the cube are generated.
Hence, \eqref{E:H3d0_F} is supplemented with two additional equations:
\begin{subequations}
\label{E:h3d0_x}
%
%
  \begin{equation}
   \label{E:h3d0_L}
p(xx_3 + x_2x_{23}) - q(xx_2 + x_3x_{23}) = 0,
  \end{equation}
  \begin{equation}
   \label{E:h3d0_Bo}
p(xx_1 + x_3x_{13}) - q(xx_3 + x_1x_{13}) = 0,
  \end{equation}
\end{subequations}
which yield solutions for $x_{12}$, $x_{13}$, and $x_{23}$:
\begin{subequations}
 \label{E:h3d0_2sol}
  \begin{equation}
   \label{E:h3d0_12Sol}
x_{12} = \frac{x(px_1 - qx_2)}{qx_1 - px_2},
  \end{equation}
  \begin{equation}
   \label{E:h3d0_13Sol} 
x_{13} = \frac{x(px_1 - kx_3)}{kx_1 - px_3}, 
  \end{equation}
  \begin{equation}
   \label{E:h3d0_23Sol}
x_{23} = \frac{x(qx_2 - kx_3)}{kx_2 - qx_3}.
  \end{equation}
\end{subequations}
Equations for the remaining faces (i.e., {\em{back, right}} and
{\em{top}}) are then generated:
\begin{subequations}
 \label{E:h3d0_x123}
  \begin{equation}
   \label{E:h3d0_Ba}
p(x_3x_{13} + x_{23}x_{123}) - q(x_3x_{23} + x_{13}x_{123}) = 0,
  \end{equation}
  \begin{equation}
   \label{E:h3d0_R}
p(x_1x_{13} + x_{12}x_{123}) - q(x_1x_{12} + x_{13}x_{123}) = 0,
  \end{equation}
  \begin{equation}
   \label{E:h3d0_T}
p(x_2x_{12} + x_{23}x_{123}) - q(x_2x_{23} + x_{12}x_{123}) = 0.
  \end{equation}
\end{subequations}
Each of these reference $x_{123}$ and thus yield three distinct
solutions for $x_{123}$, 
\begin{subequations}
 \label{E:h3d0_3sol}
 \begin{equation}
  \label{E:h3d0_123_B}
x_{123} = \frac{x_3(px_{13} - qx_{23})}{qx_{13} - px_{23}}, 
 \end{equation}
 \begin{equation}
  \label{E:h3d0_123_R}
x_{123} = \frac{x_2(px_{12} - kx_{23})}{kx_{12} - px_{23}},
 \end{equation}
 \begin{equation}
  \label{E:h3d0_123_T}
x_{123} = \frac{x_1(qx_{12} - kx_{13})}{kx_{12} - qx_{13}}. 
 \end{equation}
\end{subequations}
Remarkably, after substitution of \eqref{E:h3d0_2sol} into 
\eqref{E:h3d0_3sol} one arrives at the {\it same expression} for
$x_{123},$ namely,
%
%
\begin{equation}
 \label{E:123sol}
 x_{123} = -\frac{px_2x_3(k^2-q^2)+qx_1x_3(p^2-k^2)+kx_1x_2(q^2-p^2)}
                {px_1(k^2-q^2)+qx_2(p^2-k^2)+kx_3(q^2-p^2)} .
\end{equation}
Thus, \eqref{E:H3d0_F} is consistent around the cube. 
The consistency is apparent from the following symmetry of the right hand 
side of \eqref{E:123sol}. 
If we replace the lattice parameters $(p, q, k)$ by $(l_1,l_2,l_3)$ 
the expression would be invariant under any permutation of the 
indices $\{1,2,3\}.$

Additionally, \eqref{E:123sol} does not reference $x$.
This independence is referred to as the {\em{tetrahedron property}}.
Indeed, through \eqref{E:123sol}, the top of a tetrahedron
(located at $x_{123})$ is connected to the base of the tetrahedron 
with corners at $x_1$, $x_2$ and $x_3.$ 
\subsection{Computation of Lax pairs for scalar \PDeltaEs}
\label{sec:scLaxPair}
In analogy with the definition of Lax pairs (in matrix form) for PDEs, 
a Lax pair for a \PDeltaE\, is a pair of matrices, $(L, M),$ such that the
compatibility of the linear equations, 
for an auxiliary vector function $\psi$,
\begin{subequations}
 \begin{equation}
 \label{LP:Lmatrix}
 \psi_1 = L \, \psi , 
 \end{equation}
 \begin{equation}
 \label{LP:Mmatrix}
 \psi_2 = M \psi, 
 \end{equation}
\end{subequations}
is equivalent to the \PDeltaE. 
The crux is to find suitable matrices $L$ and $M$ so that the nonlinear 
\PDeltaE\, can be replaced by \eqref{LP:Lmatrix}-\eqref{LP:Mmatrix}. 
To avoid trivial cases, the compatibility of \eqref{LP:Lmatrix} 
and \eqref{LP:Mmatrix} should only hold on solutions of the given 
nonlinear \PDeltaE.  
%
%
%
\vskip 1pt

\noindent
\begin{figure*}[h]
 \centering
  \includegraphics[scale=1.50]{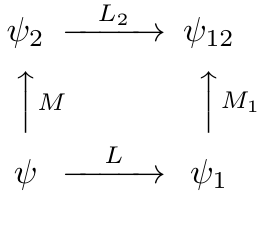}
%
  \caption{Commuting scheme resulting in the Lax equation.
  $M_1$ denotes the shift of $M$ in the $1-$direction (horizontally).
  $L_2$ denotes the shift of $L$ in the $2-$direction (vertically).}
  \label{fig:commdiag}       
\end{figure*}
\vskip 1pt
\indent
The compatibility of \eqref{LP:Lmatrix} and \eqref{LP:Mmatrix} can be 
readily expressed as follows. 
Shift \eqref{LP:Lmatrix} in the 2-direction, 
i.e., $\psi_{12} = L_2 \psi_2 = L_2 M \psi.$
Shift \eqref{LP:Mmatrix} in the 1-direction, 
i.e., $\psi_{21} = \psi_{12} = M_1 \psi_1 = M_1 L \psi,$ 
and equate the results.
Hence, $L_2 M \psi = M_1 L \psi$ must hold on solutions of the \PDeltaE.
The compatibility is visualized in Figure~\ref{fig:commdiag}, where 
commutation of the scheme indeed requires that $L_2 M = M_1 L.$
The corresponding Lax equation is thus
\begin{equation}
 \label{E:lax} 
L_2 M - M_1 L \;\;\dot{=}\;\; 0, 
\end{equation}
where $\dot{=}$ denotes that the equation holds for solutions of the \PDeltaE.

As is the case for completely integrable PDEs, Lax pairs of \PDeltaEs\, are 
not unique for they are equivalent under gauge transformations.
Specifically, if $(L, M)$ is a Lax pair then so is 
$(\mathcal{L}, \mathcal{M})$ where
\begin{equation}
\label{E:gauge}
\mathcal{L} = \mathcal{G}_1 L \mathcal{G}^{-1}, \quad
\mathcal{M} = \mathcal{G}_2 M \mathcal{G}^{-1},
\end{equation}
for any arbitrary non-singular matrix $\mathcal{G}.$
Indeed, $(\mathcal{L}, \mathcal{M})$ satisfy 
$\mathcal{L}_2 \mathcal{M} - \mathcal{M}_1 \mathcal{L} \;\;\dot{=}\;\; 0,$ 
which follows from \eqref{E:lax} by pre-multiplication by 
$\mathcal{G}_{12}$ and post-multiplication by $\mathcal{G}^{-1}.$
Alternatively, $\phi_1 = \mathcal{L} \phi$ and $\phi_2 = \mathcal{M} \phi,$
provided $\phi = \mathcal{G} \psi$.
The Lax pairs $(L, M)$ and $(\mathcal{L}, \mathcal{M})$ are said to be 
{\it gauge equivalent}.

Returning to Example 1, we show that the 
CAC property implicitly determines the Lax pair of a \PDeltaE.
%
Indeed, observe that, as a consequence of the multi-affine structure of the 
original \PDeltaE, the numerator and denominator of $x_{13}$ in 
\eqref{E:h3d0_13Sol} are linear in $x_3.$
%
%
In analogy with the linearization of Riccati equations, substitute 
$x_3 = \frac{f}{F}$ into \eqref{E:h3d0_13Sol}, yielding
\begin{equation} 
 \label{h3d0:fF13}
 x_{13} = \frac{f_1}{F_1} = \frac{fkx - Fpxx_1}{fp - Fkx_1}.
\end{equation} 
Hence, 
\begin{equation}
\label{E:f1mKdV}
f_1 = t (f k x - F p x x_1)
\end{equation} 
and 
\begin{equation}
\label{E:F1mKdV}
F_1 = t (f p - F k x_1),
\end{equation} 
where $t(x,x_1;p,k)$ is a function still to be determined.
Defining $\psi = \begin{bmatrix} f\\ F\end{bmatrix}$, system 
\eqref{E:f1mKdV}-\eqref{E:F1mKdV} can be written in matrix form
\eqref{LP:Lmatrix} where $L = t L_c$ and the ``core" of the Lax matrix 
$L$ is given by 
\begin{equation}
 \label{h3d0:Lcore}
L_c = \begin{bmatrix} 
k x & -p x x_1\\ p & -k x_1
\end{bmatrix}.
\end{equation}
%
Using \eqref{E:h3d0_23Sol}, the computation of the Lax matrix $M$ proceeds 
analogously. 
Indeed, 
\begin{equation} 
 \label{h3d0:fF23}
x_{23} = \frac{f_2}{F_2} = \frac{fkx - Fqxx_2}{fq - Fkx_2}
\end{equation} 
holds if $f_2 = s (f k x - F q x x_2)$ and $F_2 = s (f q - F k x_2)$ 
where $s(x,x_2;q,k)$ is a common factor to be determined.
Thus, we obtain \eqref{LP:Mmatrix} where $M = s M_c$ with
%
%
\begin{equation}
 \label{h3d0:Mcore}
M_c = \begin{bmatrix} 
kx & -qxx_2 \\ q & -kx_2 
\end{bmatrix}.
\end{equation}
Note that $x_{23}$ can be obtained from $x_{13}$, and hence $M_c$ from $L_c$, 
by replacing $x_1 \rightarrow x_2$ (or simply, $1 \rightarrow 2)$ and 
$p \rightarrow q.$ 
The final step is to compute $s$ and $t.$
\subsection{Determination of the scalar factors for scalar \PDeltaEs}
\label{sec:scScalarFactorsSingle}
Specific values for $s$ and $t$ can be computed using \eqref{E:lax}.
Substituting $L = t\, L_c$ and $M = s\, M_c$ yields 
%
\begin{equation}
  \label{h3d0:laxST1}
  s \, t_2  (L_c)_2 M_c - t \, s_1 (M_c)_1 L_c \;\;\dot{=}\;\; 0. 
\end{equation}
All elements in the matrix on the left hand side must vanish.
Remarkably, this yields a unique expression for the ratio
$\frac{s \, t_2}{t \, s_1}.$
\vskip 3pt
\indent
For Example 1, using \eqref{h3d0:Lcore} and \eqref{h3d0:Mcore}, 
eq.\ \eqref{h3d0:laxST1} reduces to 
\begin{equation}
 \label{h3d0:MatrixResult}
 \big( \frac{ x x_1 t \, s_1 - x x_2 s \, t_2}{p x_2 - q x_1} \big)
 \begin{bmatrix} 
 (k^2-p^2) q x_1 - (k^2-q^2) p x_2 & k (p^2 -q^2) x_1 x_2
 \\ -k (p^2 -q^2)  & (k^2-p^2) q x_1 - (k^2-q^2) p x_2 
\end{bmatrix} 
= 
\begin{bmatrix} 
0 & 0 \\ 0 & 0
\end{bmatrix}.
\end{equation}
This requires that 
\begin{equation}
 \label{h3d0:stRatio}
 \frac{s \, t_2}{t \, s_1} \;\;\dot{=}\;\;
\frac{x_1}{x_2},
\end{equation}
which has an infinite family of solutions. 
Indeed, the left hand side of \eqref{h3d0:stRatio} is invariant under the 
change 
\begin{equation}
 \label{h3d0:stChange}
t \rightarrow \frac{a_1}{a} \, t, \quad s \rightarrow \frac{a_2}{a} \, s,
\end{equation}
where $a(x)$ is arbitrary. 
%
Consistent with the notations in Section~\ref{sec:scalarPDE}, $a_1$ and $a_2$ 
denote the shifts of $a$ in the $1-$ and $2-$direction, respectively.
By inspection,  
\begin{equation}
\label{h3d0:stadd1}
t = s = \frac{1}{x} 
\end{equation}
and 
\begin{equation}
 \label{h3d0:stadd2}
t = \frac{1}{x_1}, \quad s = \frac{1}{x_2} 
\end{equation}
both satisfy \eqref{h3d0:stRatio}. 
Note that \eqref{h3d0:stadd1} can be mapped into \eqref{h3d0:stadd2} by 
taking $a = 1/x.$

Avoiding guess work, $t$ and $s$ can be computed by taking 
the determinant of \eqref{h3d0:laxST1}.
If $L_c$ and $M_c$ are $n \times n$ matrices, then 
\begin{equation}
  \label{h3d0:laxST2}
  ( s \, t_2)^n \text{det}\, (L_c)_2 \, \text{det}\, M_c 
  = (t \, s_1)^n \text{det}\, (M_c)_1 \, \text{det}\, L_c, 
\end{equation}
yielding
\begin{equation}
\label{h3d0:stRat}
  \frac{s \, t_2}{t \, s_1} 
  = \sqrt[n]{\frac{\text{det} (M_c)_1 \, \text{det} L_c}
            {\text{det} (L_c)_2 \, \text{det} M_c}}, 
\end{equation}
which is satisfied by 
\begin{equation}
 \label{h3d0:stvals}
t = \frac{1}{\sqrt[n]{\text{det} L_c}}, \quad
s = \frac{1}{\sqrt[n]{\text{det} M_c}}.
\end{equation}
For Example 1, i.e., eq.\ \eqref{E:H3d0_F}, by substituting  
\eqref{h3d0:Lcore} and \eqref{h3d0:Mcore} into \eqref{h3d0:stRat}, 
one then obtains
 \begin{equation}
  \label{h3d0:sVal}
  t = \frac{1}{\sqrt{(p^2-k^2) x x_1}},
  \quad
  s = \frac{1}{\sqrt{(q^2-k^2) x x_2}}.
\end{equation}
The constant factors involving $p, q$ and $k$ are irrelevant.
Therefore, \eqref{h3d0:sVal} can be replaced by 
 \begin{equation}
  \label{h3d0:sValsimple}
  t = \frac{1}{\sqrt{x x_1}},
  \quad
  s = \frac{1}{\sqrt{x x_2}}.
\end{equation}
Thus, using the {\it determinant method}, a Lax pair for \eqref{E:H3d0_F} is 
\begin{equation}
  \label{h3d0:LaxLM} 
L = \frac{1}{\sqrt{x x_1}}
    \begin{bmatrix} 
    k x & -p x x_1 \\ p & -k x_1
    \end{bmatrix}, \quad 
M = \frac{1}{\sqrt{x x_2}}
    \begin{bmatrix} 
    k x & -q x x_2 \\ q & -k x_2 
    \end{bmatrix}.
\end{equation}
The irrational $t$ and $s$ in \eqref{h3d0:sValsimple} can be transformed into 
\eqref{h3d0:stadd1}, by taking $a = \sqrt{x}$, or 
into \eqref{h3d0:stadd2}, by $a = \frac{1}{\sqrt{x}},$ 
both yielding rational Lax pairs.

\section{Systems of partial difference equations}
\label{sec:systemPDE}
Section~\ref{sec:scalarPDE} dealt with single (scalar) \PDeltaEs, 
i.e., equations involving only one field variable (denoted by $x$).
This section covers systems of \PDeltaEs\, defined on quadrilaterals
involving multiple field variables. 
Here we will consider examples involving three field variables 
$x, y,$ and $z.$
Figures~\ref{fig:square} and~\ref{fig:cube} still apply provided we 
replace the scalar $x$ by vector ${\bf x} \equiv (x,y,z).$ 
Hence, 
${\bf x}_1 = (x_1, y_1, z_1), \, {\bf x}_2 = (x_2, y_2, z_2), \,
{\bf x}_{12} = ( x_{12}, y_{12}, z_{12} ),$ etc.
\subsection{Consistency around the cube for systems of \PDeltaEs}
\label{sec:sysCAC}
To apply the algorithm in Section \ref{sec:scLaxPair} to
systems of \PDeltaEs, it is necessary to maintain consistency for all 
equations on all six faces of the cube, handle the edge equations in an 
appropriate way, and ultimately arrive at the same expressions for 
$x_{123}$, as well as for $y_{123}$ and $z_{123}.$ 
%
%
%
\vskip 3pt
\noindent
{\bf Example 2}:
Consider the lattice Schwarzian Boussinesq system \cite{Nijhoff1996}:
\begin{subequations}
 \label{E:bsq}
 \begin{equation}
  \label{E:bsq1} 
x_1 y - z_1 + z = 0, 
 \end{equation}
 \begin{equation}
  \label{E:bsq2} 
x_2 y - z_2 + z = 0, 
 \end{equation}
 \begin{equation}
  \label{E:bsq3}
x y_{12} (y_1 - y_2) - y (p x_1 y_2 - q x_2 y_1) = 0.
 \end{equation}
\end{subequations}
Eqs.\ \eqref{E:bsq1} and \eqref{E:bsq2} are defined along a single edge 
of the square while \eqref{E:bsq3} is defined on the whole square.
The edge equations, unlike the face equation, can be shifted in the 
1- or 2-directions while still remaining on the square.
Then, \eqref{E:bsq} is augmented with additional shifted edge equations, 
\begin{subequations}
 \label{E:bsq-edge}
 \begin{equation}
 \label{E:bsq4}
x_{12} y_2 - z_{12} + z_2 = 0,
 \end{equation}
 \begin{equation}
 \label{E:bsq5}
x_{12} y_1 - z_{12} + z_1 = 0, 
 \end{equation}
\end{subequations}
obtained from \eqref{E:bsq1} and \eqref{E:bsq2}, respectively.
Solving for the variables ${\bf x}_{12} = (x_{12}, y_{12}, z_{12})$ 
referenced in the augmented system 
(i.e., \eqref{E:bsq} augmented with \eqref{E:bsq-edge}) gives 
\begin{subequations}
 \label{E:bsq-12sol}
 \begin{equation}
  \label{E:bsq-x12}
x_{12} = \frac{z_2 - z_1}{y_1 - y_2},
 \end{equation}
 \begin{equation}
  \label{E:bsq-y12}
y_{12} = \frac{y(px_1y_2 - qx_2y_1)}{x(y_1 - y_2)}, 
 \end{equation}
 \begin{equation}
  \label{E:bsq-z12}
z_{12} = \frac{y_1z_2 - y_2z_1}{y_1 - y_2}.
 \end{equation}
\end{subequations}
Continuing as before by generating the variations of \eqref{E:bsq} on the 
faces of the cube and solving for the variables with double subscripts 
yields ${\bf x}_{13}$ and ${\bf x}_{23}$.
Indeed, from the equations on the {\textit{bottom}} face (not shown)
%
%
one gets ${\bf x}_{13}$ with components 
\begin{subequations}
 \label{E:bsq-13sol}
 \begin{equation}
  \label{E:bsq-x13}
x_{13} = \frac{z_3 - z_1}{y_1 - y_3}, 
 \end{equation}
 \begin{equation}
  \label{E:bsq-y13}
y_{13} = \frac{y(px_1y_3 - kx_3y_1)}{x(y_1 - y_3)}, 
 \end{equation}
 \begin{equation}
  \label{E:bsq-z13}
z_{13} = \frac{y_1z_3 - y_3z_1}{y_1 - y_3},
 \end{equation}
\end{subequations}
which readily follow from \eqref{E:bsq-12sol} by replacing 
${\bf x}_2 \rightarrow {\bf x}_3, {\bf x}_{12} \rightarrow{\bf x}_{13},$ 
and $q\rightarrow k$.
Or simpler, $2 \rightarrow 3$ and $q \rightarrow k.$
Similarly, the equations on the {\textit{left}} face of the cube
%
determine ${\bf x}_{23}$ with components
\begin{subequations}
 \label{E:bsq-23sol}
 \begin{equation}
  \label{E:bsq-x23}
x_{23} = \frac{z_2 - z_3}{y_3 - y_2}, 
 \end{equation}
 \begin{equation}
  \label{E:bsq-y23}
y_{23} = \frac{y(kx_3y_2 - qx_2y_3)}{x(y_3 - y_2)}, 
 \end{equation}
 \begin{equation}
  \label{E:bsq-z23}
z_{23} = \frac{y_3z_2 - y_2z_3}{y_3 - y_2},
 \end{equation}
\end{subequations}
easily obtained 
by a change of labels and parameters, namely,
$1 \rightarrow 2, p \rightarrow q, 2 \rightarrow 3,$ and $q \rightarrow k).$ 
%
Likewise, the equations on the {\textit{back}} face (not shown) 
%
%
determine ${\bf x}_{123}$ with components
\begin{subequations}
 \label{E:bsq-123solB}
 \begin{equation}
  \label{E:bsq-x123b}
x_{123} = \frac{z_{23} - z_{13}}{y_{13} - y_{23}},
 \end{equation}
 \begin{equation}
  \label{E:bsq-y123b}
y_{123} = \frac{y_3(px_{13}y_{23}-qx_{23}y_{13})}{x_3(y_{13}-y_{23})}, 
 \end{equation}
 \begin{equation}
  \label{E:bsq-z123b}
z_{123} = \frac{y_{13}z_{23} - y_{23}z_{13}}{y_{13} - y_{23}},
 \end{equation}
\end{subequations}
which follow from \eqref{E:bsq-12sol} by applying the shift in the 
third direction, which amounts to ``adding" a label 3 to all variables. 
%
Similarly, the equations on the {\textit{right}} face (suppressed)
%
yield ${\bf x}_{123}$ with components
\begin{subequations}
 \label{E:bsq-123solR}
 \begin{equation}
  \label{E:bsq-x123r}
x_{123} = \frac{z_{12} - z_{13}}{y_{13} - y_{12}}, 
 \end{equation}
 \begin{equation}
  \label{E:bsq-y123r}
y_{123} = \frac{y_1(kx_{13}y_{12} - qx_{12}y_{13})}
                 {x_1(y_{13} - y_{12})},
 \end{equation}
 \begin{equation}
  \label{E:bsq-z123r}
z_{123} = \frac{y_{13}z_{23} - y_{12}z_{13}}{y_{13} - y_{12}},
 \end{equation}
\end{subequations}
which follow from \eqref{E:bsq-23sol} by applying a shift in the 1-direction.
Finally, the equations on the {\textit{top}} face (suppressed)
%
yield 
\begin{subequations}
 \label{E:bsq-123solT}
 \begin{equation}
  \label{E:bsq-x123t}
x_{123} = \frac{z_{23} - z_{12}}{y_{12} - y_{23}}, 
 \end{equation}
 \begin{equation}
  \label{E:bsq-y123t}
y_{123} = \frac{y_2(p x_{12} y_{23} - k x_{23} y_{12})}
                 {x_2 (y_{12} - y_{23})}, 
 \end{equation}
 \begin{equation}
  \label{E:bsq-z123t}
z_{123} = \frac{y_{12} z_{23} - y_{23} z_{12}}{y_{12} - y_{23}} ,
 \end{equation}
\end{subequations}
obtained from \eqref{E:bsq-13sol} by a shift in the 2-direction.

%
Using \eqref{E:bsq-12sol}-\eqref{E:bsq-23sol} to evaluate the expressions 
\eqref{E:bsq-123solB}-\eqref{E:bsq-123solT} yields the 
{\it same} ${\bf x}_{123}$ with
\begin{subequations}
 \label{E:bsq123}
 \begin{equation}
  \label{E:bsqx123}
x_{123} = \frac{x (x_1 - x_2) \big( y_1 (z_2 - z_3) + y_2 (z_3 - z_1) + 
                  y_3 (z_1 - z_2) \big)}
               {(z_1 - z_2) \big( p x_1 (y_3 - y_2) + q x_2 (y_1 - y_3) + 
                 k x_3 (y_2 - y_1) \big)} ,
 \end{equation}
 \begin{equation}
  \label{E:bsqy123}
y_{123} = \frac{q (z_2 - z_1) (k x_3 y_1 - p x_1 y_3) + 
                k (z_3 - z_1) (p x_1 y_2 - q x_2 y_1)} 
               {x_1 \big( p x_1 (y_3 - y_2) + q x_2 (y_1 - y_3) + 
                k x_3 (y_2 - y_1) \big)}, 
%
 \end{equation}
 \begin{equation}
  \label{E:bsqz123}
z_{123} = 
\frac{p x_1 (y_3 z_2 - y_2 z_3) + q x_2 (y_1 z_3 - y_3 z_1)
      + k x_3 (y_2 z_1 - y_1 z_2) }
      { p x_1 (y_3 - y_2) + q x_2 (y_1 - y_3) + k x_3 (y_2 - y_1) }.
%
 \end{equation}
\end{subequations}
Thus, \eqref{E:bsq} is {\it multi-dimensionally} consistent around the cube,
i.e., the systems of \PDeltaEs\ is consistent around the cube with respect 
to each component of ${\bf x},$ i.e., $x$, $y$ and $z$.

The expressions for $x_{123}$ and $y_{123}$ can be written in more symmetric 
form by eliminating $z_1, z_2,$ and $z_3.$
To do so, we use the edge equations
\begin{subequations}
\label{E:bsqL}
\begin{equation}
\label{E:bsqL1} 
x_3 y - z_3 + z = 0, 
\end{equation}
\begin{equation}
\label{E:bsqL2} 
x_2 y - z_2 + z = 0, 
\end{equation}
\end{subequations}
defined on the {\textit{left}} face of the cube.
Subtracting \eqref{E:bsq1} from \eqref{E:bsq2} and \eqref{E:bsqL1} from 
\eqref{E:bsqL2} yields
\begin{equation}
\label{E:bsq-fraczx}
\frac{z_2 - z_1}{x_2 - x_1} 
= \frac{z_3 - z_2}{x_3 - x_2} 
= \frac{z_3 -z_1}{x_3 - x_1}
= y.
\end{equation}
%
%
%
Using the above ratios, \eqref{E:bsqx123} and \eqref{E:bsqy123} can be 
replaced by 
\begin{subequations}
 \label{E:bsq123alt}
 \begin{equation}
  \label{E:bsqx123alt}
x_{123} = 
   \frac{ x \Big( y_1 (x_2 - x_3) + y_2 (x_3 - x_1) + y_3 (x_1 - x_2) \Big) }
        {p x_1 (y_3 - y_2) + q x_2 (y_1 - y_3) + k x_3 (y_2 - y_1)} ,
%
%
%
 \end{equation}
 \begin{equation}
  \label{E:bsqy123alt}
y_{123} = \frac{y \Big(k q y_1 (x_2 - x_3) + k p y_2 (x_3 - x_1) +
                 p q y_3(x_1 - x_2) \Big) } 
               {p x_1 (y_3 - y_2) + q x_2 (y_1 - y_3) + 
                 k x_3 (y_2 - y_1)}, 
 \end{equation}
%
%
\end{subequations}
Before continuing with the calculations of a Lax pair, 
%
%
it is worth noting that \eqref{E:bsq} does not satisfy the tetrahedron 
property because $x$ explicitly appears in the right hand side of 
\eqref{E:bsqx123}.
The impact of not having the tetrahedron property remains unclear but 
does not affect the computation of a Lax pair.
%
%
\subsection{Computation of a Lax pair for systems of \PDeltaEs}
\label{sec:sysLaxPair}
Both the numerators and denominators of the components of ${\bf x}_{13}$ and 
${\bf x}_{23}$ (in \eqref{E:bsq-13sol} and \eqref{E:bsq-23sol}, 
respectively), are affine linear in the components of ${\bf x}$. 
Due to their linearity in $x_3$, $y_3$ and $z_3$, substitution of fractional 
expressions for $x_3$, $y_3$ and $z_3$ will allow one to compute Lax
matrices.
%
%
In contrast to the scalar case, the computations are more subtle because 
the edge equations on the left face of the cube 
introduce constraints between $x_3$ and $z_3$.

%
Continuing with Example 2, solving \eqref{E:bsqL1} for $x_3$ yields
\begin{equation}
\label{E:bsq-x3sol}
x_3 = \frac{z_3 - z}{y}.
\end{equation}
%
Therefore, setting 
\begin{subequations}
 \label{E:bsq-subs}
 \begin{equation}
  \label{E:bsq-subsZ}
z_3 = \frac{f}{F}
 \end{equation}
and
 \begin{equation}
  \label{E:bsq-subsY}
y_3 = \frac{g}{G}
 \end{equation}
determines 
 \begin{equation}
  \label{E:bsq-subsX}
x_3 = \frac{z_3 - z}{y} = \frac{f - Fz}{Fy}.
 \end{equation}
\end{subequations}
Substituting \eqref{E:bsq-subs} into \eqref{E:bsq-13sol} then yields
\begin{subequations}
 \label{E:bsq-subxyz}
 \begin{equation}
  \label{E:bsq-subsX13}
x_{13} = \frac{G(F z_1 - f)}{F (g - G y_1)}, 
 \end{equation}
 \begin{equation}
  \label{E:bsq-subsY13}
y_{13} = \frac{Gfky_1 - Fgx_1y - FGky_1z}{Fx(g - Gy_1)}, 
 \end{equation}
 \begin{equation}
  \label{E:bsq-subsZ13}
z_{13} = \frac{Fgz_1 - Gfy_1}{F(g - Gy_1)},
 \end{equation}
\end{subequations}
%
which are not yet linear in $f$, $g$, $F$ and $G.$
Additional constraints between $f$, $g$, $F$ and $G$ will achieve this goal.
Indeed, setting $G = F$ 
%
%
simplifies \eqref{E:bsq-subxyz} into
\begin{subequations}
 \label{E:bsq-subxyzF}
 \begin{equation}
  \label{E:bsq-subFx13}
x_{13} = \frac{f - F z_1}{F y_1 - g},
 \end{equation}
 \begin{equation}
  \label{E:bsq-subFy13}
y_{13} = \frac{g p x_1 y - f k y_1 + F k y_1 z}{x (F y_1 - g)},
 \end{equation}
 \begin{equation}
  \label{E:bsq-subFz13}
z_{13} = \frac{f y_1 - g z_1}{F y_1 - g} .
 \end{equation}
\end{subequations}
Simultaneously, \eqref{E:bsq-subs} reduces to 
\begin{subequations}
 \label{E:bsq-finalSub}
 \begin{equation}
  \label{E:bsq-finalSubZ}
z_3 = \frac{f}{F},
 \end{equation}
 \begin{equation}
  \label{E:bsq-finalSubY}
y_3 = \frac{g}{F}, 
 \end{equation}
 \begin{equation}
  \label{E:bsq-finalSubX}
x_3 = \frac{f - Fz}{Fy}, 
 \end{equation}
\end{subequations}
whose shifts in the 1-direction must be compatible with 
\eqref{E:bsq-subxyzF}.
Equating $z_{13} = \frac{f_1}{F_1}$ to \eqref{E:bsq-subFz13} requires that 
\begin{equation}
 \label{E:bsq-f1}
f_1 = t\, (f y_1 - g z_1)
\end{equation} 
and 
\begin{equation}
 \label{E:bsq-F1}
F_1 = t\, (F y_1 - g).
\end{equation}  
Next, equating $y_{13} = \frac{g_1}{F_1}$ with \eqref{E:bsq-subFy13} gives 
\begin{equation}
\label{E:bsq-g1}
g_1 = t \, \frac{1}{x} (g p x_1 y - f k y_1 + F k y_1 z).
\end{equation}  
Finally, one has to verify that the 1-shift of \eqref{E:bsq-finalSubX}, 
%
%
\begin{equation}
 \label{E:bsq-x13FG1}
x_{13} = \frac{f_1 - F_1 z_1}{F_1 y_1}, 
\end{equation}
matches \eqref{E:bsq-subFx13}.
That is indeed the case.
After substitution of $f_1$ and $F_1$ into \eqref{E:bsq-x13FG1} 
\begin{equation}
 \label{E:bsq-x13FG2}
x_{13} = \frac{t (f y_1 - g z_1) - t (F y_1 - g) z_1}{t (F y_1 - g) y_1}
       = \frac{f - F z_1}{F y_1 - g}.
\end{equation}
%
Defining 
$\psi = 
\begin{bmatrix} g \\ f \\ F 
\end{bmatrix},$
eqs.\ \eqref{E:bsq-f1}-\eqref{E:bsq-g1} can be written in matrix form
%
%
yielding \eqref{LP:Lmatrix} with
\begin{equation}
 \label{E:bsq-tLcore-x3}
L = t \begin{bmatrix}
        \frac{px_1y}{x}&-\frac{ky_1}{x}&\frac{ky_1z}{x}\\
         -z_1 & y_1 &0\\ 
         -1 & 0 & y_1
       \end{bmatrix},
\end{equation}
where $t({\bf x}, {\bf x}_1; p, k).$
%
%
%
Similarly, from \eqref{E:bsq-23sol} one derives 
%
%
\begin{equation}
 \label{E:bsq-sMcore}
M = s \begin{bmatrix}
        \frac{q x_2 y}{x}&-\frac{k y_2}{x}&\frac{k y_2 z}{x}\\
         -z_2 & y_2 &0\\ 
         -1 & 0 & y_2
       \end{bmatrix}, 
\end{equation}
which can also be obtained from \eqref{E:bsq-tLcore-x3} by applying
the replacement rules $1 \rightarrow 2$ and $p \rightarrow q.$
%
%
%
\subsection{Determination of the scalar factors for systems of \PDeltaEs}
\label{sec:scScalarFactorsSystems}
As discussed in Section~\ref{sec:scScalarFactorsSingle}, specific values 
for $s$ and $t$ may be  computed algorithmically using \eqref{h3d0:stvals}.
For Example 2, this yields
%
%
\begin{equation}
 \label{E:bsq-tVal}
%
t = \frac{1}{\sqrt[3]{\frac{(k-p) y_1^2 (z - z_1)}{x}}}, \quad
%
s = \frac{1}{\sqrt[3]{\frac{(k-q) y_2^2 (z - z_2)}{x}}}.
\end{equation}
Cancelling trivial factors, a Lax pair for \eqref{E:bsq} is thus given by 
\begin{subequations}
 \label{E:bsq-LaxPair}
  \begin{equation}
   \label{E:bsq-LaxPairL}
L = \sqrt[3]{\frac{x}{y_1^2 (z - z_1)}}
    \begin{bmatrix}
     \frac{px_1y}{x}&-\frac{ky_1}{x}&\frac{ky_1z}{x}\\
     -z_1 & y_1 &0\\
     -1 & 0 & y_1
    \end{bmatrix}, 
  \end{equation}
  \begin{equation}
   \label{E:bsq-LaxPairM}
M = \sqrt[3]{\frac{x}{y_2^2 (z - z_2)}}
%
    \begin{bmatrix}
     \frac{qx_2y}{x}&-\frac{ky_2}{x}&\frac{ky_2z}{x}\\
     -z_2 & y_2 &0\\ 
     -1 & 0 & y_2
    \end{bmatrix}.
  \end{equation}
\end{subequations}
Unfortunately, these matrices have irrational functional factors. 
Using \eqref{E:lax} we find the following equation for the scalar factors
%
\begin{equation}
 \label{E:bsq-stRatio}
\frac{s \, t_2}{t \, s_1} \;\; \dot{=} \;\; \frac{y_1}{y_2}.
\end{equation}
%
%
Once can easily verify that \eqref{E:bsq-stRatio} is satisfied by
\begin{equation}
\label{E:bsq-stadd}
t = s = \frac{1}{y} \;\; \text{and}\;\; 
t \;\; = \frac{1}{y_1},\; s = \frac{1}{y_2},
\end{equation}
which both yield rational Lax pairs. 
The factors $t, s$ in \eqref{E:bsq-stadd} are related to those in 
\eqref{E:bsq-tVal}.
Using \eqref{E:bsq1}, $t$ in \eqref{E:bsq-tVal} can be written as
\begin{equation}
 \label{E:bsq-tValnew}
t = \sqrt[3]{\frac{x}{(p-k) y_1^2 y x_1}} .
%
\end{equation}
After applying \eqref{h3d0:stChange} with $a = \sqrt[3]{x/y},$ 
one can simplify the cube root to find $t = 1/{y_1},$ 
where the trivial factor $1/\sqrt[3]{p-k}$ has been canceled.
A further application of \eqref{h3d0:stChange} with $a = y$ then yields 
$t = 1/y.$ 
The connections between the choices for $s$ are similar.

An alternate form of a Lax pair is possible.
Had the original constraint given by \eqref{E:bsqL1} been expressed as
\begin{equation}
 \label{E:bsq-z3sol}
z_3 = x_3y + z,
\end{equation}
the substitutions would become
%
\begin{subequations}
 \label{E:bsq-subsbis}
 \begin{equation}
  \label{E:bsq-subsZbis}
x_3 = \frac{{\tilde{f}}}{{\tilde{F}}}, 
 \end{equation}
 \begin{equation}
  \label{E:bsq-subsYbis}
y_3 = \frac{{\tilde{g}}}{{\tilde{F}}}, 
 \end{equation}
 \begin{equation}
  \label{E:bsq-subsXbis}
z_3 = \frac{{\tilde{f}} y + {\tilde{F}} z}{{\tilde{F}}}.
 \end{equation}
\end{subequations}
With 
$\phi = 
\begin{bmatrix} 
{\tilde{f}} \\ {\tilde{g}} \\ {\tilde{F}} 
\end{bmatrix},$ 
$\mathcal{L}$ would then be given by 
\begin{equation}
 \label{E:bsq-tLcore-z3}
\mathcal{L} = t\begin{bmatrix}
      \frac{px_1y}{x} & -\frac{kyy_1}{x} & 0\\ 
      0 & y & z - z_1\\
      -1 & 0 & y_1 
     \end{bmatrix}.
\end{equation}
Note that the matrices \eqref{E:bsq-tLcore-x3} and \eqref{E:bsq-tLcore-z3} 
are gauge equivalent as defined in \eqref{E:gauge} with 
\begin{equation}
 \label{E:bsq-gauge}
\mathcal{G} = 
\begin{bmatrix} 
1 & 0 & 0 \\ 
0 & 1/y & -z/y \\ 
0 & 0 & 1 \end{bmatrix}.
\end{equation}
\section{Implementation}
\label{sec:implement}
\subsection{Consistency around the Cube}
\label{sec:implementCAC}
The CAC property has been used to identify integrable \PDeltaEs\,
\cite{AdlerBobenkoSuris2002,Hietarinta2011}.
As shown in both examples, the information gained from the process of 
verifying CAC is also crucial to the computation of the corresponding
Lax pair. 
In some sense the lattice equation is its own Lax pair, cf.\ the discussion 
in \cite{Nijhoff2002}.

For scalar \PDeltaEs, CAC is a simple concept that can be verified by 
hand or (interactively) with a computer algebra system (CAS) such as 
{\sc Mathematica} or {\sc Maple}.
Hereman \cite{Hereman2009} designed software to compute Lax pairs of 
scalar ${\rm P}\Delta{\rm Es}$ defined on a quadrilateral.
For systems of \PDeltaEs\, with edge equations the verification of the CAC
property can be tricky and the order in which substitutions are carried
out is important.
Designing a symbolic manipulation package that fully automates the steps is 
quite a challenge \cite{Bridgman2012}.

Naively, one could first generate the comprehensive system that represents 
the \PDeltaEs\, on each face of the cube and then ask a CAS to solve it.
To be consistent around the cube, that system should have a unique solution 
for ${\bf x}_{123}.$ 
%
Wolf \cite{Wolf2008} discusses the computational challenges of verifying the 
CAC property for scalar \PDeltaEs\ in 3 dimensions \cite{TsarevWolf2008}
due to the astronomical size of the overdetermined system that has to be 
solved.
Even for \PDeltaEs\, in 2 dimensions, in particular, those involving edge 
equations, automatically solving such a system often exceeds the capabilities 
of current symbolic software packages.
It is therefore necessary to verify CAC in a more systematic way like one 
would do with pen on paper. 
%
%

Computer code \cite{Bridgman2012} for automated verification of the CAC 
property carries out the following steps:
\begin{enumerate}
 \item Solve the initial \PDeltaE\, for ${\bf x}_{12}.$ 
 Solve the equations on the \textit{bottom} and \textit{left} faces for 
 ${\bf x}_{13}$ and ${\bf x}_{23}$, respectively.
 Generate the equations for the \textit{back}, \textit{right}
 and \textit{top} equations
 and solve each for ${\bf x}_{123}.$ 
 This produces three expressions for the components of ${\bf x}_{123}.$

 \item Evaluate and simplify the solutions ${\bf x}_{123}$ using 
 ${\bf x}_{12}, {\bf x}_{13},$ and ${\bf x}_{23}.$ 
 Use the constraints between the components of 
 ${\bf x}, {\bf x}_1, {\bf x}_2,$ and ${\bf x}_3$ arising from the edge
 equations to check consistency at every level of the computation.
 
 \item Finally, verify if the three expressions for the 
 components of ${\bf x}_{123}$ are indeed equal. 
 If so, the system of \PDeltaEs\, is consistent around the cube and
 one can proceed with the computation of the Lax matrices.

\end{enumerate} 
\subsection{Computation of a Lax pair}
\label{sec:compLax}
Assuming the given \PDeltaE\, is CAC, the following steps are then taken 
to calculate a Lax pair:
\begin{enumerate}
 \item Introduce fractional expressions (e.g., $\frac{f}{F}, \frac{g}{G},$ 
  etc.) for the various components of ${\bf x}_3$
  in order to linearize the numerators and denominators of the 
  expressions for ${\bf x}_{13}$ in terms of $f, F, g, G,$ etc.

 \item Further simplify the components of ${\bf x}_3$ 
  using the edge equations (if present in the given \PDeltaE).

 \item Substitute the simplified expressions for ${\bf x}_3$ 
  into ${\bf x}_{13}$
  and again examine if the numerators and denominators are linear in 
  $f, F, g, G,$ etc.

 \item If ${\bf x}_{13}$ is not yet ``linearized", reduce the degree 
  of freedom (e.g., by setting $G = F,$ etc.) and repeat this procedure
  until the numerators and denominators of the components of ${\bf x}_{13}$ 
  are linear in $f, F, g,$ etc.

 \item Use the fractional linear expressions of ${\bf x}_{13}$ to generate
  the ``core" Lax matrix $L_c.$ 

 \item Use the determinant method (see \eqref{h3d0:stvals}) to compute 
 a possible scaling factor $t.$ 

 \item The Lax matrix is then $L = t L_c.$ 
 The matrix $M = s M_c$ follows from $L$ by replacing $p$ by $q$ 
 and ${\bf x}_1$ by ${\bf x}_2.$ 
\end{enumerate}

\subsection{Verification of the Lax pair}
\label{sec:verify}
Finally, verify the Lax pair by substitution into the Lax equation 
\eqref{E:lax}.
Unfortunately, the determinant method gives $s$ and $t$ in irrational form,
introducing, e.g., square or cubic roots into the symbolic computations.
In general, symbolic software is limited in simplification of expressions 
involving radicals.
%
%
The impact of the presence of radical expressions can be reduced by careful 
simplification. 
Notice that
\eqref{h3d0:laxST1} can be written as
\begin{subequations}
 \begin{equation}
 \label{E:bsq-laxVerify1}
\frac{(s \, t_2)}{(t\, s_1)}(L_c)_2 M_c - (M_c)_1 L_c \;\;\dot{=}\;\; 0.
 \end{equation}
Bringing all common factors from the matrix products up front gives
 \begin{equation}
 \label{E:bsq-laxVerify2}
\left(\frac{s \, t_2}{t\, s_1}\frac{{\rm CF}_{L_2M}}{{\rm CF}_{M_1L}}\right)
  \tilde{L}_2 \tilde{M} - \tilde{M}_1\tilde{L} \;\;\dot{=}\;\; 0 
 \end{equation}
where ${\rm CF}_{X}$ stands for a common factor of all the entries of 
a matrix $X$.
Hence, $\dsp {\rm CF}_{L_2M} \, \tilde{L}_2 \tilde{M} = (L_c)_2 M_c$
and $\dsp {\rm CF}_{M_1L} \, \tilde{M}_1 \tilde{L} = (M_c)_1 L_c.$
\end{subequations}
The computed Lax pair is correct if
\begin{subequations}
 \label{E:bsq-VerCond}
 \begin{equation}
\label{E:st2ts1trick}
\left(\frac{s\, t_2}{t\, s_1}\frac{{\rm CF}_{L_2M}}{{\rm CF}_{M_1L}}\right)
  \;\; \dot{=}\;\; \pm 1\\
 \end{equation}
and, thus
 \begin{equation}
\label{E:Laxeqtrick}
\pm \tilde{L_2}\tilde{M} - \tilde{M_1}\tilde{L} \;\;\dot{=}\;\; 0.
 \end{equation}
\end{subequations}
%
%
To illustrate the verification procedure, consider Example 2 with 
$t$ and $s$ in \eqref{E:bsq-tVal}.
Here, 
\begin{subequations}
\begin{equation}
\label{E:bsq-ratioexplicit}
 \frac{s\, t_2}{t\, s_1} = 
  \frac{ \sqrt[3]{\frac{x}{(k - q) y_2^2 (z - z_2)}}
         \sqrt[3]{\frac{x^2(y_2-y_1)^3(z - z_2)}
         {(k - p)yy_2(py_2(z_1-z) + qy_1(z-z_2))^2(z_1-z_2)}}}
       {\sqrt[3]{\frac{x}{(k - p) y_1^2 (z - z_1)}}
         \sqrt[3]{\frac{x^2(y_2-y_1)^3(z - z_1)}
        {(k - q)yy_1(py_2(z_1-z) + qy_1(z-z_2))^2(z_1-z_2)}}}, 
\end{equation}
\begin{equation}
\label{E:bsq-commonfactors}
{\rm CF}_{L_2M} = \frac{y_2}{x(y_1 - y_2)} \; \text{ and } \;
{\rm CF}_{M_1L} = \frac{y_1}{x(y_1 - y_2)}.
\end{equation}
\end{subequations}
%

The matrix $\tilde{L_2}\tilde{M}$ (which equals $\tilde{M_1}\tilde{L})$ is 
\begin{equation}
 \begin{bmatrix}
-pqy(z_1-z_2)   & ky(qy_1 - py_2)          & ky (py_2z_1 - qy_1z_2) \\
& & \\
 pz_2(z-z_1)+qz_1(z_2-z) & k(y_1z_2-y_2z_1)         & kz(y_2z_1 - y_1z_2) \\
                         & + \,p y_2(z_1-z)+qy_1(z -z_2) & \\
& & \\
  p (z-z_1) + q (z_2-z) & k(y_1 - y_2)       &  kz(y_2-y_1)+ py_2(z_1-z) \\
   &                          & +\, qy_1(z-z_2) 
 \end{bmatrix}.
\end{equation}
Note that 
\begin{equation}
\label{E:bsq-CFfraction}
\frac{{\rm CF}_{L_2M}}{{\rm CF}_{M_1L}} = \frac{y_2}{y_1}.
\end{equation}
After multiplying \eqref{E:bsq-CFfraction} with \eqref{E:bsq-ratioexplicit}, 
the resulting expression can be simplified\footnote{Use the 
{\sc Mathematica} function {\sc PowerExpand} or simply cube the expression.} 
into 1. 
Thus, both \eqref{E:st2ts1trick} and \eqref{E:Laxeqtrick} are satisfied for 
the plus sign. 
%
%
\section{Results}
\label{sec:results}
The algorithm discussed in this paper is being implemented in 
{\sc Mathematica} and preliminary versions of the software 
\cite{Bridgman2012,Hereman2009} are being verified against many known 
\PDeltaEs.
The Lax matrices $L$, including those for Examples 1 and 2 in the paper, 
are presented in Tables~\ref{table:scalarA} through~\ref{table:system5}.
The matrix $M$ follows from the matrix $L$ by the replacements 
${\bf x}_1 \rightarrow {\bf x}_2$ and $p \rightarrow q$. 
%
\subsection{Scalar \PDeltaEs}
\label{results:scalar}
The scalar \PDeltaEs\, given in Tables~\ref{table:scalarA} 
and~\ref{table:scalarB} are referenced by the names given in the 
classification by Adler, Bobenko, and Suris \cite{AdlerBobenkoSuris2002}. 
Each of these \PDeltaEs\, involves the scalar field variable $x$ and 
its shifts. 
The substitution used in the computation of a Lax pair is 
\begin{equation}
 \label{scalar:sub}
 x_3 = \frac{f}{F} .
\end{equation}
Thus, the linear equations have the form 
\eqref{LP:Lmatrix}-\eqref{LP:Mmatrix}, in which
\begin{equation}
 \label{scalar:psi}
 \psi = 
 \begin{bmatrix} 
 f \\ F 
 \end{bmatrix}.
\end{equation}
Scaling factors can be computed with the determinant method but they are 
often irrational. 
%
%
%
If for scalar \PDeltaEs\, the ratio $\frac{s \, t_2}{t \, s_1}$ can 
be factored, i.e.,
\begin{equation}
 \label{E:ratiofactors}
 \frac{s \, t_2 }{t \, s_1} = 
  \frac{\mathcal{P}(x, x_1; p, q) \, \mathcal{Q}(x, x_1; p, q)}
       {\mathcal{P}(x, x_2; q, p) \, \mathcal{Q}(x, x_2; q, p)},
\end{equation}
then potential candidates for the scaling factors are
\begin{equation}
 \label{E:ratioCand}
 t = \frac{1}{\mathcal{P}(x, x_1; p, q)}, \;\;
 s = \frac{1}{\mathcal{P}(x, x_2; q, p)}
 \;\; \text{and} \;\;
 t = \frac{1}{\mathcal{Q}(x, x_1; p, q)}, \;\;
 s = \frac{1}{\mathcal{Q}(x, x_2; q, p)}. 
\end{equation}
To verify that the candidate scaling factors actually work, 
$L = t\, L_c$ and $M = s\, M_c$ must satisfy \eqref{E:lax}. 
If they do work, such $t$ and $s$ are rational and preferred over the 
irrational scaling factors computed by the determinant method.
%
%
The alternative rational scaling factors, obtained in this way, 
are listed for $Q_1$ and the $(\alpha, \beta)$-equation
in Table~\ref{table:scalarB}.
The Lax pair for the $(\alpha, \beta)$-equation was first presented in 
\cite{Tran2007}. 

A similar situation happens with $Q_3$ when $\delta = 0$ where
in addition to the irrational expression of $t$ one has two rational 
alternatives, namely, $t = 1/(p x - x_1)$ and $t = 1/(p x_1 - x)$
which both satisfy
\begin{equation}
\label{Q3d0:stRatio}
\frac{s \, t_2}{t \, s_1} \;\; \dot{=} \;\;
\frac{(q^2 - 1)(p x - x_1)(p x_1 - x)}{(p^2 -1)(q x - x_2)(q x_2 - x)}.
\end{equation}

For the equations $A_1$ and $A_2$ in Table~\ref{table:scalarA},  
the ratio $\frac{s \, t_2}{t \, s_1}$ is also of the form 
\eqref{E:ratiofactors} but the choices \eqref{E:ratioCand} are not valid.
The irrational forms of $t$ and $s$ as listed in Table~\ref{table:scalarA}
have to be used. 

The Lax pair for Example 1, i.e., \eqref{E:H3d0_F}, follows from the one 
for $H_3$ by setting $\delta = 0.$ 
However, when $\delta = 0$, the factors $t$ and $s$ can be taken 
rational (see \eqref{h3d0:stadd1} and \eqref{h3d0:stadd2}).

Further alternate rational factors are obtained using \eqref{h3d0:stChange}
for the Schwarzian, modified, Toda-modified Boussinesq equations as well 
as the Hietarinta systems.
%
%
%
%
\subsection{Systems of \PDeltaEs}
\label{results:systems}
\subsubsection{Boussinesq Systems}
\label{sys:BSQ}
For the Boussinesq system \cite{Nijhoffetal1992} in Table~\ref{table:system3}, 
$\psi = 
\begin{bmatrix} F \\ f \\ g 
\end{bmatrix}.$
Substitution of 
\begin{equation}
 \label{bsq:subs-x}
  x_3 = \frac{f}{F},\quad
  y_3 = \frac{g}{F}, \;\; \text{and} \;\;
  z_3 = \frac{fx - Fy}{F}, 
\end{equation}
yields the Lax matrix given in Table~\ref{table:system3}. 
 
Representing the edge constraint as $x_3 = \frac{z_3 + y}{x}$ requires
\begin{equation}
 \label{bsq:subs-y}
  x_3 = \frac{\tf + \tF y}{\tF x},\quad
  y_3 = \frac{\tg}{\tF}, \;\; \text{and} \;\;
  z_3 = \frac{\tf}{\tF}. 
\end{equation}
For $\phi = 
\begin{bmatrix} \tF \\ \tf \\ \tg 
\end{bmatrix},$
a resulting gauge equivalent $\mathcal{L}$ matrix is then
\begin{equation}
 \label{bsq:Ly}
  \mathcal{L} = \frac{1}{x}\begin{bmatrix}
       xx_1 - y & -1 & 0\\
       yy_1 & y_1 & -xx_1\\
       x(k - p + xy_1) - z(xx_1 - y) & z & -x^2
      \end{bmatrix}, 
\end{equation}
%
where the gauge matrix, cf.\ \eqref{E:gauge}, is given by 
\begin{equation}
 \label{bsq:G}
  \mathcal{G} = \begin{bmatrix}
       1 & 0 & 0\\
       y/x & 1/x & 0\\ 
       0 & 0 & 1
      \end{bmatrix}. 
\end{equation}
\subsubsection{Hietarinta Systems}
\label{sys:hietarinta}
For each system given in Table~\ref{table:system4},
$\psi = 
\begin{bmatrix} f \\ g \\ G 
\end{bmatrix}.$
However, the substitutions are impacted by the edge equations in the systems.
For system A-2, the edge constraint was represented as 
$x_3 = \frac{x+y_3}{z}$ resulting in substitutions of 
\begin{equation}
 \label{hietA-2:subs-x}
  x_3 = \frac{g+Gx}{Gz},\quad y_3=\frac{g}{G}, \;\; \text{and} \;\;
  z_3 = \frac{f}{G}. 
\end{equation}
Writing the edge constraint as $y_3 = x + x_3 z$ requires one to work with 
\begin{equation}
 \label{hietA-2:subs-y}
  x_3 = \frac{\tg}{\tG},\quad
  y_3 = \frac{\tG x - \tg z}{\tG}, \;\; \text{and} \;\;
  z_3 = \frac{\tf}{\tG}. 
\end{equation}
Setting 
$\phi = 
\begin{bmatrix} \tf \\ \tg \\ \tG 
\end{bmatrix},$
the resulting gauge equivalent $\mathcal{L}$ matrix is given by
\begin{equation}
 \label{HietA-2:Ly}
  \mathcal{L} = \begin{bmatrix}
        \frac{y}{x} & \frac{k}{x} & -\frac{px_1+yz_1}{x}\\
        0 & 1 & -x_1\\ 1 & 0 & -z_1
       \end{bmatrix} , 
\end{equation}
where $L$ and $\mathcal{L}$ are connected as shown in \eqref{E:gauge} with 
\begin{equation}
 \label{HietA-2:G}
  \mathcal{G} = \begin{bmatrix}
       1 & 0 & 0\\
       0 & 1/z & x/z\\
       0 & 0 & 1
      \end{bmatrix}.
\end{equation}
For system B-2, the edge constraint was represented as 
$x_3 = \frac{z+y_3}{x}$ resulting in 
\begin{equation}
 \label{hietB-2:subs-x}
  x_3 = \frac{g+Gz}{Gx},\quad 
  y_3 = \frac{g}{G}, \;\; \text{and} \;\;
  z_3 = \frac{f}{G}. 
\end{equation}
Representing the edge constraint as $y_3 = z + x_3 x$ yields
\begin{equation}
 \label{hietB-2:subs-y}
  x_3 = \frac{\tg}{\tG},\quad 
  y_3 = \frac{\tg x -\tG z}{\tG}, \;\; \text{and} \;\;
  z_3 = \frac{\tf}{\tG}. 
\end{equation}
With 
$\phi = 
\begin{bmatrix} 
\tf \\ \tg \\ \tG 
\end{bmatrix}$ 
the resulting gauge equivalent $\mathcal{L}$ matrix is given by
\begin{equation}
 \label{HietB-2:Ly}
  \mathcal{L} = \begin{bmatrix}
        \delta +x & -(x\delta+y) & k-p+x_1(x\delta+y) - z_1(\delta+x)\\
        1 & 0 & -z_1\\ 0 & 1 & -x_1
       \end{bmatrix},
\end{equation}
%
where $L$ and $\mathcal{L}$ are connected (cf.\ \eqref{E:gauge}) by 
\begin{equation}
 \label{HietB-2:G}
  \mathcal{G} = \begin{bmatrix}
       1 & 0 & 0\\
       0 & 1/x & z/x\\ 
       0 & 0 & 1
      \end{bmatrix}. 
\end{equation}
For system C-3, the edge constraint was represented as $x_3 = x + zy_3$ and 
\begin{equation}
 \label{hietC-3:subs-x}
  x_3 =\frac{Gx+gz}{Gx},\quad y_3=\frac{g}{G}, \;\; \text{and} \;\;
  z_3 =\frac{f}{G}. 
\end{equation}
Representing the edge constraint as $y_3 = \frac{x_3 - x}{z}$ requires 
\begin{equation}
 \label{hietC-3:subs-y}
  x_3 = \frac{\tg}{\tG},\quad 
  y_3 = \frac{\tg - \tG x}{\tG z}, \;\; \text{and} \;\;
  z_3 = \frac{\tf}{\tG}. 
\end{equation}
Letting 
$\phi = 
\begin{bmatrix} \tf \\ \tg \\ \tG
\end{bmatrix},$
a gauge equivalent $\mathcal{L}$ matrix is 
\begin{equation}
 \label{HietC-3:Ly}
  \mathcal{L} = \frac{1}{z}\begin{bmatrix}
    \frac{\delta_1+x\delta_2-pzy_1}{y} & \frac{kz_1}{y} & 
           -\frac{(\delta_1+x\delta_2+kx)z_1}{y}\\
       x_1 & -z_1 & 0\\ 1 & 0 & -z_1
      \end{bmatrix}, 
\end{equation}
%
with gauge matrix  
\begin{equation}
 \label{HietC-3:G}
  \mathcal{G} = \begin{bmatrix}
       1 & 0 & 0\\
       0 & z & x\\ 0 & 0 & 1
      \end{bmatrix}. 
\end{equation}
For system C-4, the edge constraint was represented as $x_3 = x + z y_3.$ 
Hence,
\begin{equation}
 \label{hietC-4:subs-x}
  x_3 = \frac{Gx+gz}{Gx},\quad
  y_3 = \frac{g}{G}, \;\; \text{and} \;\;
  z_3 = \frac{f}{G}. 
\end{equation}
Representing the edge constraint as $y_3 = \frac{x_3 - x}{z}$ requires
\begin{equation}
 \label{hietC-4:subs-y}
  x_3 = \frac{\tg}{\tG},\quad
  y_3 = \frac{\tg - \tG x}{\tG z}, \;\; \text{and} \;\;
  z_3 = \frac{\tf}{\tG}. 
\end{equation}
With 
$\phi = 
\begin{bmatrix} 
\tf \\ \tg \\ \tG 
\end{bmatrix},$ 
a resulting gauge equivalent $\mathcal{L}$ matrix is 
\begin{equation}
 \label{HietC-4:Ly}
  \mathcal{L} = \frac{1}{z}\begin{bmatrix}
       \frac{\delta_1+xx_1-pzy_1}{y} & \frac{(k-x)z_1}{y} & 
           -\frac{(\delta_1+kx)z_1}{y}\\
       x_1 & -z_1 & 0\\ 1 & 0 & -z_1
      \end{bmatrix}, 
\end{equation}
%
with gauge matrix
\begin{equation}
 \label{HietC-4:G}
  \mathcal{G} = \begin{bmatrix}
       1 & 0 & 0\\
       0 & z & x\\ 0 & 0 & 1
      \end{bmatrix}. 
\end{equation}
\subsubsection{Two-component pKdV and NLS lattices}
\label{sys:pKdV}
In finding a Lax pair for the two-component pKdV system \cite{Xenitidis2009}
given in Table~\ref{table:system5}, the initial substitutions are
\begin{equation}
 \label{pKdV:subs}
  x_3 = \frac{f}{F} \;\; \text{and}\ ;\; y_3 = \frac{g}{G}, 
\end{equation}
which lead to the proper form of the components of ${\bf x}_{13}.$
Thus, the resulting Lax pair comprises $4 \times 4$ matrices as the 
linear equations involve the auxiliary vector
\begin{equation}
 \label{pkdv:psi}
 \psi = 
 \begin{bmatrix} f \\ F \\ g \\ G 
 \end{bmatrix}.
\end{equation}
Also, an additional scaling factor is introduced by the disparate 
substitutions.
In this case, the constraints on the scaling factors become
\begin{equation}
 \label{pKdV:tCons}
  t \,T = \frac{1}{\sqrt{\text{det} \, L_c}} = \frac{1}{p-k}.
\end{equation}
Hence, one can take $t = T = 1.$

For the lattice NLS system \cite{Xenitidis2009} given in 
Table~\ref{table:system5}, one is only able to solve for $x_{13}$ and 
$x_{23}$ despite having equations referencing $y.$
Thus, the substitution of 
$ x_3 = \frac{f}{F} $
suffices to linearize the components of ${\bf x}_{13}.$
The resulting Lax matrices, $L$ and $M,$ are $2 \times 2$ matrices and 
\begin{equation}
 \label{nls:psi}
  \psi = 
  \begin{bmatrix} f \\ F 
  \end{bmatrix}.
\end{equation}
%
%
\section{Conclusion}
\label{sec:concl}
We gave a detailed review of a three-step method 
\cite{BobenkoSuris2002,Nijhoff2002} to compute Lax pairs for scalar 
${\rm P}\Delta{\rm Es}$ defined on quadrilaterals and subsequently 
applied the method to systems of ${\rm P}\Delta{\rm Es}.$ 
It was shown that for systems involving edge equations the derivation of 
Lax pairs can be quite tricky.

The paper also serves as a repository of Lax pairs, not only for the scalar 
integrable ${\rm P}\Delta{\rm Es}$ classified by Adler, Bobenko, and Suris 
\cite{AdlerBobenkoSuris2002}, but for systems of ${\rm P}\Delta{\rm Es}$ 
including the discrete potential KdV equation, as well as various nonlinear 
Schr\"odinger and Boussinesq-type lattices. 
Previously unknown Lax pairs are presented for ${\rm P}\Delta{\rm Es}$
recently derived by Hietarinta \cite{Hietarinta2011}.

Preliminary software \cite{Hereman2009} is available to compute Lax pairs of 
scalar ${\rm P}\Delta{\rm Es}$ defined on quadrilaterals.
The extension of the code to systems of ${\rm P}\Delta{\rm Es}$ is a 
nontrivial exercise.
In the near future we hope to release a fully-automated {\sc Mathematica} 
package \cite{Bridgman2012} for the computation (and verification) of 
Lax pairs of two-dimensional ${\rm P}\Delta{\rm Es}$ systems defined 
on quadrilaterals.
%
%
\section{Acknowledgments}
\label{sec:ackno}
%
%
The research is supported in part by the Australian Research Council (ARC) 
and the National Science Foundation (NSF) of the U.S.A.\ under Grant 
No.\ CCF-0830783.
Any opinions, findings, and conclusions or recommendations expressed in 
this material are those of the authors and do not necessarily reflect the 
views of ARC or NSF.

WH is grateful for the hospitality and support of the Department of 
Mathematics and Statistics of La Trobe University (Melbourne, Australia) 
where this project was started in November 2007.

The authors thank the Isaac Newton Institute for Mathematical Sciences 
(Cambridge, U.K.) where the work was continued during the 
{\it Programme on Discrete Integrable Systems} in Spring 2009.
\vfill
\newpage
%
%

\vfill
\newpage
\newpage
\thispagestyle{plain}

\begin{sidewaystable}
\caption{Lax pairs of scalar \PDeltaEs}
\vspace{2mm}
%
%
\centering
\begin{tabular}{|p{1cm}|l|p{9cm}|p{3cm}|p{5mm}|}
 \hline
  & & & & \\
 Name & Equation & Matrix $L$ & Alternate $t$ values & Ref.\ \\
  & & & & \\ \hline
  & & & & \\
 $H_1$
 \label{ABS:H1} & 
 \parbox{6.5cm}{
  \begin{align*}
   (x - x_{12})(x_1 - x_2) + q - p = 0
  \end{align*}
 } & 
 \parbox{7.5cm}{ 
  $\dsp t\begin{bmatrix}
  x & p - k - xx_1 \\ 
  1 & -x_1
  \end{bmatrix}$
  with $\dsp t = 1$
 } &  & \cite{AdlerBobenkoSuris2002} \\
  & & & & \\ \hline
  & & & & \\
 $H_2$ 
 \label{ABS:H2} & 
 \parbox{6.5cm}{
   \begin{align*}
    (&x - x_{12})(x_1 - x_2) \\
     &+ (q - p)(x + x_1+x_2+x_{12}+p+q)  = 0
   \end{align*}
 } & 
 \parbox{7.5cm}{ 
  $\dsp t\begin{bmatrix}
   p - k + x & p^2 - k^2 + (p-k)(x+x_1) - xx_1 \\ 
   1 & -(p - k + x_1)
  \end{bmatrix}$ \\\\\\
  with $\dsp t = \frac{1}{\sqrt{p + x + x_1}}$
 } & & \cite{AdlerBobenkoSuris2002} \\
  & & & & \\ \hline
  & & & & \\
 $H_3$ 
 \label{ABS:H3} & 
 \parbox{6.5cm}{
  \begin{align*}
    p&(xx_1 + x_2x_{12}) - q(xx_2 + x_1x_{12}) \\
      &+ \delta(p^2 - q^2)  = 0
  \end{align*} 
 } &
 \parbox{7.5cm}{  
  $\dsp t\begin{bmatrix}
    kx & -\delta(p^2 - k^2) - pxx_1\\ 
    p & -kx_1
   \end{bmatrix}$ \\\\\\
  with $\dsp t = \frac{1}{\sqrt{\delta p + x x_1}}$
 } & & \cite{AdlerBobenkoSuris2002} \\
  & & & & \\ \hline
  & & & & \\
 $A_1$ 
 \label{ABS:A1} & 
 \parbox{6.5cm}{
  \begin{align*}
    p&(x + x_2)(x_1+x_{12}) - q(x + x_1)(x_2+x_{12})\\
     &- \delta^2pq(p - q)  = 0
  \end{align*} 
 } &
 \parbox{7.5cm}{  
  $\dsp t\begin{bmatrix}
    kx + \eta x_1 & -p(xx_1+\delta^2k\eta)\\ 
    p & -(kx_1+\eta x)
   \end{bmatrix}$
  where $\dsp \eta = k - p$, \\\\\\
  with 
$\dsp t = \frac{1} { \sqrt{ (\delta p - (x+x_1))(\delta p + (x+x_1)) }}$
 } & & \cite{AdlerBobenkoSuris2002} \\
  & & & & \\ \hline
  & & & & \\
 $A_2$ 
 \label{ABS:A2} & 
 \parbox{6,5cm}{
  \begin{align*}
    (&q^2-p^2)(xx_1x_2x_{12}+1) \\
       &+ q(p^2-1)(xx_2+x_1x_{12}) \\
       &- p(q^2-1)(xx_1 + x_2x_{12})  = 0
  \end{align*} 
 }&
 \parbox{7.5cm}{  
  $\dsp t\begin{bmatrix}
    k\gamma x & -(\tau+p\sigma xx_1)\\ 
    p\sigma+\tau xx_1 & -k\gamma x_1
   \end{bmatrix}$ \\\\ 
   where
   \begin{equation*}
    \gamma = p^2-1,\hspace{3mm} \sigma = k^2 - 1 \text{ and }
    \tau = p^2 - k^2
   \end{equation*}\\
with $\dsp t = \frac{1}{\sqrt{-(p-xx_1)(pxx_1-1)}}$
  } & & \cite{AdlerBobenkoSuris2002} \\
  & & & & \\ \hline
\end{tabular}
\label{table:scalarA}
\end{sidewaystable}
%

\newpage

\begin{sidewaystable}
\caption{Lax pairs of scalar \PDeltaEs\, -- continued}
\vspace{2mm}
%
%
\centering
\begin{tabular}{|p{1cm}|l|p{8cm}|p{4cm}|p{5mm}|}
 \hline
  & & & & \\
 Name & Equation & Matrix $L$ & Alternate $t$ values & Ref.\ \\
  & & & & \\ \hline
  & & & & \\
 $Q_1$ 
 \label{ABS:Q1} & 
 \parbox{6.5cm}{
  \begin{align*}
    p&(x - x_2)(x_1 - x_{12}) - q(x - x_1)(x_2 - x_{12}) \\
     &+ \delta^2pq(p - q)  = 0
  \end{align*} 
 } & 
 \parbox{7.5cm}{  
  $\dsp t\begin{bmatrix}
    kx-\eta x_1 & -p(xx_1-\delta^2k\eta)\\ 
    p & -(kx_1-\eta x)
   \end{bmatrix}$
  where $\dsp \eta = k - p$, \\\\\\
  with 
$\dsp t = \frac{1} {\sqrt{-(\delta p - (x-x_1))(\delta p + (x-x_1))}}$
 } &
 \parbox{7.5cm}{  
  $\dsp t = \frac{1}{\delta p \pm(x-x_1)}$
 } & \cite{AdlerBobenkoSuris2002} \\ 
  & & & & \\ \hline
 $Q_2$ 
 \label{ABS:Q2} &
 \parbox{6.5cm}{
  \begin{align*}
    p&(x - x_2)(x_1 - x_{12}) - q(x - x_1)(x_2 - x_{12})\\
     &+ pq(p - q)(x + x_1 + x_2 + x_{12})\\
     &- pq(p - q)(p^2 - pq + q^2) = 0
  \end{align*}
  } & 
  \parbox{7.5cm}{ 
   $\dsp t\begin{bmatrix}
    \eta(kp-x_1)+kx & \ell_{12}\\ 
    p & -(\eta(kp-x)+kx_1
   \end{bmatrix}$ \\\\ 
   where
   \begin{equation*}
     \ell_{12} = -p[k\eta(k\eta+p^2-x-x_1)+xx_1], \text{ and}
   \end{equation*}
   \begin{equation*}
    \eta = k - p,\\
   \end{equation*}
with $\dsp t = \frac{1}{\sqrt{ (x-x_1)^2-2 p^2(x+x_1)+p^4 }}$
  } & & \cite{AdlerBobenkoSuris2002} \\
  & & & &  \\ \hline
  & & & & \\
 $Q_3$ 
 \label{ABS:Q3} &
 \parbox{6.5cm}{
  \begin{align*}
    (&q^2 - p^2)(xx_{12} + x_1x_2) \\
     &+ q(p^2 - 1)(xx_1 + x_2x_{12})\\
     &- p(q^2 - 1)(xx_2 + x_1x_{12}) \\
     &- \frac{\delta^2}{4pq}(p^2 - q^2)(p^2 - 1)(q^2 - 1) = 0
  \end{align*}
  } & 
  \parbox{7.5cm}{ 
   $\dsp t\begin{bmatrix}
    -4kp(\sigma px+\tau x_1) & -\gamma(\delta^2\sigma\tau - 
                                                4k^2pxx_1)\\ 
    -4k^2p\gamma & 4kp(\sigma px_1+\tau x)
   \end{bmatrix}$ \\\\ 
   where
   \begin{equation*}
    \gamma = p^2-1,\hspace{3mm}\sigma = k^2 - 1 \text{ and }
    \tau = p^2 - k^2,
   \end{equation*}
    with\\ 
$\dsp t = \frac{1}
{\sqrt{4p (px - x_1)(px_1 - x) - \delta^2\gamma^2}}$
  } & 
  \parbox{4.5cm}{ when $\delta=0$,
  \begin{align*}
   \!\!\!\!\!\!\!\!\!\!\!\!\!\!\!\!\!\!\!\!\!\!t &= \frac{1}{p x - x_1}, 
   \; \text{or} \\
   \!\!\!\!\!\!\!\!\!\!\!\!\!\!\!\!\!\!\!\!\!\!t &= \frac{1}{p x_1 - x}
  \end{align*}
  }
  & \cite{AdlerBobenkoSuris2002} \\
  & & & &  \\ \hline
  & & & & \\
 $(\alpha, \beta)-{\rm lattice}$ 
 \label{alphaBeta} &
 \parbox{6.5cm}{
  \begin{align*}
    (&(p - \alpha)x - (p + \beta)x_1)\\
     &\times ((p - \beta)x_2 - (p + \alpha)x_{12})\\
     & - ((q - \alpha)x - (q + \beta)x_2)\\
     & \times((q - \beta)x_1 - (q + \alpha)x_{12}) = 0
  \end{align*}
 } & 
  \parbox{7.5cm}{ 
   $\dsp t\begin{bmatrix}
    (p-\alpha)(p-\beta)x - \tau x_1 & -(k-\alpha)(k-\beta)xx_1\\ 
    (k+\alpha)(k+\beta) & -((p+\alpha)(p+\beta)x_1 - \tau x)
   \end{bmatrix} \;$ \\\\ 
   where
   \begin{equation*}
    \tau = p^2 - k^2
   \end{equation*}
   with\\ $\dsp t = \frac{1}{\sqrt{((\beta-p)x+(\alpha+p)x_1)
                                    ((\alpha-p)x+(\beta+p)x_1)}}$
 } &
 \parbox{3.5cm}{  
  \begin{align*}
   t &= \frac{1}{(\alpha-p)x+(\beta+p)x_1}, \\
     & \text{or} \\
   t &= \frac{1}{(\beta-p)x+(\alpha+p)x_1}
  \end{align*}
 } &  
   \parbox{0.20cm}{
   \begin{align*}
   \cite{NijhoffQuispelCapel1983} \\
   \cite{Tran2007}
   \end{align*}
 } \\
  & & & &  \\ \hline
\end{tabular}
\label{table:scalarB}
\end{sidewaystable}
%

\begin{sidewaystable}
\caption{Lax pairs of systems of \PDeltaEs}
\vspace{2mm}
%
%

\centering
\begin{tabular}{|p{2cm}|l|p{8cm}|p{3cm}|p{5mm}|}
 \hline
  & & & & \\
 Name & Equation & Matrix $L$ & Alternate $t$ values & Ref.\ \\
  & & & & \\ \hline
  & & & & \\
 Boussinesq &
 \parbox{6.5cm}{
  \begin{align*}
   &z_1 - xx_1 + y = 0\\
   &z_2 - xx_2 + y = 0\\
   &(x_2 - x_1)(z - xx_{12}+y_{12})-p+q = 0
  \end{align*}
 } & 
 \parbox{7.5cm}{ 
  $\dsp t 
  \begin{bmatrix}
    -x_1 & 1 & 0\\ 
    -y_1&0&1\\
    p-k-xy_1 + x_1z&-z&x
   \end{bmatrix}$
   with $\dsp t = 1$
 } &  & \cite{Nijhoffetal1992}\\
  & & & & \\ \hline
  & & & & \\
 Schwarzian Boussinesq &
 \parbox{6.5cm}{ 
  \begin{align*}
    &z_1 - yx_1 - z = 0\\
    &z_2 - yx_2 - z = 0\\
    &xy_{12}(y_1 - y_2)- y(px_1y_2 - qx_2y_1) = 0
 \end{align*} 
 } & 
 \parbox{7.5cm}{ 
  $\dsp t\begin{bmatrix}
    \frac{pyx_1}{x}&-\frac{ky_1}{x}&\frac{kzy_1}{x}\\
    -z_1 & y_1 & 0\\ 
    -1 & 0 & y_1
   \end{bmatrix}$ \\\\\\ 
   with $\dsp t = \sqrt[3]{\frac{x}{y_1^2(z_1 - z)}}$
 } &
 $\dsp \begin{aligned}
   t &= \frac{1}{y}, \;\text{or} \\ 
   t &= \frac{1}{y_1}
 \end{aligned}$ & \cite{Nijhoff1996}\\
  & & & & \\ \hline
  & & & & \\
 Modified Boussinesq &
 \parbox{6.5cm}{
  \begin{align*}
   &x_{12}(py_1 - qy_2) - y(px_2 - qx_1) = 0\\
   &xy_{12}(py_1 - qy_2) - y(px_1y_2 - qx_2y_1) = 0
  \end{align*}
 } & 
  \parbox{7.5cm}{ 
   $\dsp t\begin{bmatrix}
     py_1 & 0 & -k\\ 
     -kx_1y & py & 0\\
     0 & -\frac{kyy_1}{x} & \frac{px_1y}{x}
    \end{bmatrix}$ \\\\\\
     with $\dsp t = \sqrt[3]{\frac{x}{x_1y^2y_1}}$ 
  } &
 $\dsp \begin{aligned}
   t &= \frac{1}{y}, \;\text{or} \\ 
   t &= \frac{1}{y_1}
 \end{aligned}$ & \cite{XenitidisNijhoff2011}\\
  & & & & \\ \hline
  & & & & \\
 \parbox{1.5cm}{Toda- \\ modified\\ Boussinesq} &
 \parbox{6.5cm}{
  \begin{align*}
   &y_{12}(p-q + x_2 - x_1) \\
    &- (p-1)y_2 + (q-1)y_1 = 0\\
   &y_1y_2(p-q-z_2+z-1) \\
    &-(p-1)yy_2 + (q-1)yy_1 = 0\\
  & y(p+q -z - x_{12})(p-q+x-2-x_1) \\
    &-(p^2+p+1)y_1 + (q^2+q+1)y_2 = 0
  \end{align*}
 } & 
  \parbox{7.5cm}{ 
 $\dsp t\begin{bmatrix}
  k+p-z&\frac{1+k+k^2}{y}&\ell_{13}\\ 
  0&p-1&(1-k)y_1\\
  1&0&p-k-x_1
  \end{bmatrix}$ \\\\ 
  where
  \begin{equation*}
  \begin{split}
   \ell_{13} = (p^2 - k^2) - x_1(p + k) &+ z(k-p+x_1) \\
                      &- \frac{y_1}{y}(p^2+p+1)
   \end{split}
  \end{equation*}
   with $\dsp t = \sqrt[3]{\frac{y}{y_1}}$
  } &
 $\dsp t = 1$ & \cite{Nijhoffetal1992}\\
  & & & & \\ \hline
\end{tabular}
\label{table:system3}
\end{sidewaystable}
%
%

\begin{sidewaystable}
\caption{Lax pairs of systems of \PDeltaEs\, -- continued}
\vspace{2mm}
%
%
\centering
\begin{tabular}{|p{2cm}|l|p{8cm}|p{3cm}|p{5mm}|}
 \hline
  & & & & \\
 Name & Equation & Matrix $L$ & Alternate $t$ values & Ref.\ \\
  & & & & \\ \hline
  & & & & \\
  A-2 &
 \parbox{6.5cm}{
  \begin{align*}
   &x_1z - y_1 - x = 0\\
   &x_2z - y_2 - x = 0\\
   &z_{12} - \frac{y}{x} - 
   \frac{1}{x}\left(\frac{px_1 - qx_2}{z_1 - z_2}\right) = 0
  \end{align*}
 } & 
 \parbox{7.5cm}{ 
  $\dsp t\begin{bmatrix}
    \frac{yz}{x} & \frac{k}{x} & \frac{kx - px_1z - yzz_1}{x}\\ 
    -x_1z & z_1 & xz_1\\
     z & 0 & -zz_1
   \end{bmatrix}$ \\\\\\ 
   with $\dsp t = \sqrt[3]{\frac{x}{x_1z^2z_1}}$
 } &
 $\dsp \begin{aligned}
   t &= \frac{1}{z}, \;\text{or} \\ 
   t &= \frac{1}{z_1}
 \end{aligned}$ & \cite{Hietarinta2011}\\
  & & & & \\ \hline
  & & & & \\
  B-2 &
 \parbox{6.5cm}{ 
  \begin{align*}
    &x_1x - y_1 - z = 0\\
    &x_2x - y_2 - z = 0\\
    &z_{12} + y -\delta(x_{12} - x)-xx_{12} - \frac{p-q}{x_1-x_2} = 0
 \end{align*} 
 } & 
 \parbox{7.5cm}{ 
  $\dsp t\begin{bmatrix}
    -(\delta x + x^2) & \delta x + y & \ell_{13}\\ 
    -x x_2 & z_2 & z z_2\\
    0 & -1 & x x_2 - z
   \end{bmatrix}$ \\\\ 
  where
  \begin{equation*}
  \begin{split}
   \ell_{13} = (z - x x_2)(\delta x + y) + z_2(\delta x + x^2)\\
                   + x(q - k)\\
   \end{split}
  \end{equation*}
   with $\dsp t = \frac{1}{\sqrt[3]{x^2 x_1}}$
 } &
 $\dsp \begin{aligned}
   t &= \frac{1}{x}, \;\text{or} \\ 
   t &= \frac{1}{x_1}
 \end{aligned}$ & \cite{Hietarinta2011}\\
  & & & & \\ \hline
  & & & & \\
 C-3  &
 \parbox{6.5cm}{
  \begin{align*}
    &y_1z - x_1 + x = 0\\
    &y_2z - x_2 + x = 0\\
    &z_{12} - \frac{\delta_2x + \delta_1}{y} - 
      \frac{z}{y} \left( \frac{py_1z_2 - qy_2z_1}{z_1 - z_2} \right) = 0
 \end{align*} 
 } & 
  \parbox{7.5cm}{ 
 $\dsp t\begin{bmatrix}
  \frac{\delta_1 + \delta_2 x - py_1z}{y}&\frac{kzz_1}{y}&
        -\frac{\delta_1z_1+\delta_2xz_1}{y}\\ 
  0&-z&x_1 - x\\
  1&0&-z_1
  \end{bmatrix}$ \\\\\\
   with $\dsp t = \sqrt[3]{\frac{y}{y_1z^2z_1}}$
  } &
 $\dsp \begin{aligned}
   t &= \frac{1}{z}, \;\text{or} \\ 
   t &= \frac{1}{z_1}
 \end{aligned}$ &  \cite{Hietarinta2011}\\
  & & & & \\ \hline
  & & & & \\
  C-4  &
 \parbox{6.5cm}{
  \begin{align*}
    &y_1z - x_1 + x = 0\\
    &y_2z - x_2 + x = 0\\
    &z_{12} - \frac{xx_{12} - \delta_1}{y} - 
      \frac{z}{y}\left( \frac{py_1z_2 - qy_2z_1}{z_1 - z_2} \right) = 0
 \end{align*} 
 } & 
  \parbox{7.5cm}{ 
 $\dsp t\begin{bmatrix}
  \frac{\delta_1 + xx_1 - py_1z}{y}&\frac{(k - x) z z_1}{y}&
        -\frac{(\delta_1+x^2)z_1}{y}\\ 
  0&-z&x_1 - x\\
  1&0&-z_1
  \end{bmatrix}$ \\\\\\
   with $\dsp t = \sqrt[3]{\frac{y}{y_1z^2z_1}}$
  } &
 $\dsp \begin{aligned}
   t &= \frac{1}{z}, \;\text{or} \\ 
   t &= \frac{1}{z_1}
 \end{aligned}$ &  \cite{Hietarinta2011}\\
  & & & & \\ \hline
\end{tabular}
\label{table:system4}
\end{sidewaystable}
%
%
\begin{sidewaystable}
\caption{Lax pairs of systems of \PDeltaEs\, -- continued}
\vspace{2mm}
%
%
\centering
\begin{tabular}{|p{2cm}|l|p{8cm}|p{3cm}|p{5mm}|}
 \hline
  & & & & \\
 Name & Equation & Matrix $L$ & Alternate $t$ values & Ref.\ \\
  & & & & \\ \hline
  & & & & \\
 C-2.1  &
 \parbox{6.5cm}{
  \begin{align*}
    &x_{12} - \frac{x_2z_1 - x_1z_2}{z_1-z_2} = 0\\
    &z_{12} + zx_{12} - \frac{z(pz_2 - qz_1)}{z_1 - z_2} = 0
 \end{align*} 
 } & 
  \parbox{7.5cm}{ 
 $\dsp t\begin{bmatrix}
   -z_1 & x_1 & 0\\ 
   zz_1 & -z(p+x_1) & kzz_1\\
   0 & 1 & -z_1
  \end{bmatrix}$ \\\\\\
   with $\dsp t = \frac{1}{\sqrt[3]{z z_1^2}}$
  } &
 $\dsp \begin{aligned}
   t &= \frac{1}{z}, \;\text{or} \\ 
   t &= \frac{1}{z_1}
 \end{aligned}$ &  \cite{Hietarinta2011}\\
  & & & & \\ \hline
  & & & & \\
  C-2.2  &
 \parbox{6.5cm}{
  \begin{align*}
    &x_{12} - \frac{x_2z_1 - x_1z_2}{z_1-z_2} = 0\\
    &z_{12} + \delta\frac{z}{x} - 
         \frac{z}{x}\left(\frac{px_1z_2 - qx_2z_1)}{z_1 - z_2}\right) = 0
 \end{align*} 
 } & 
  \parbox{7.5cm}{ 
  $\dsp t\begin{bmatrix}
    -z_1 & x_1 & 0\\ 
    \frac{k z z_1}{x}&-\frac{z}{x}(\delta+px_1)&\frac{\delta z z_1}{x}\\
    0 & 1 & -z_1
   \end{bmatrix}$ \\\\\\
%
   with $\dsp t = \sqrt[3]{\frac{x}{z z_1^2 x_1}}$
  } &
 $\dsp \begin{aligned}
   t &= \frac{1}{z}, \text{ or} \\ 
   t &= \frac{1}{z_1}
 \end{aligned}$ &  \cite{Hietarinta2011}\\
  & & & & \\ \hline
  & & & & \\
 Two-component pKdV &
 \parbox{6.5cm}{
  \begin{align*}
   &(x - x_{12})(y_1 - y_2) - p^2 + q^2 = 0\\
   &(y - y_{12})(x_1 - x_2) - p^2 + q^2 = 0
  \end{align*}
 } & 
  \parbox{7.5cm}{ 
   $\dsp \begin{bmatrix}
     0 & 0 & tx & t(p^2 - k^2 - xy_1)\\ 
     0 & 0 & t & -ty_1\\
     Ty & T(p^2 - k^2 - x_1y) & 0 & 0\\
     T & -Tx_1 & 0 & 0
    \end{bmatrix}$ \\\\\\ 
    with $t = T = 1$ 
  } &  & 
   \parbox{0.20cm}{
   \begin{align*}
   \cite{Mikhailov2009} \\
   \cite{Xenitidis2009}
   \end{align*}
 } \\
  & & & & \\ \hline
  & & & & \\
 Lattice NLS &
 \parbox{6.5cm} {
  \begin{align*}
   &y_1 - y_2 - y((x_1 - x_2)y + p - q) = 0\\
   &x_1 - x_2 + x_{12}((x_1 - x_2)y + p - q = 0
  \end{align*}
 } & 
 $\dsp t\begin{bmatrix}
  -1 & x_1\\ 
  y & k - p - yx_1
  \end{bmatrix}$ with $\dsp t = 1$ &  & 
   \parbox{0.20cm}{
   \begin{align*}
   \cite{Mikhailov2009} \\
   \cite{Xenitidis2009}
   \end{align*}
 } \\
  & & & & \\ \hline
\end{tabular}
\label{table:system5}
\end{sidewaystable}
%
%

\end{document}